\documentclass[journal]{IEEEtran}
\usepackage{amsmath,amsfonts}
\usepackage{amssymb}
\usepackage{tabularx}
\usepackage{amsthm}  
\usepackage{subcaption}   
\usepackage{array}
\usepackage{arydshln}
\usepackage{enumitem}
\usepackage[caption=false,font=normalsize,labelfont=sf,textfont=sf]{subfig}
\usepackage{textcomp}
\usepackage{stfloats}
\usepackage{url}
\usepackage{verbatim} 
\usepackage{graphicx}
\usepackage{color}
\usepackage{booktabs}  
\usepackage[linesnumbered, ruled]{algorithm2e} 
\usepackage{graphicx}
\usepackage{pgfplots}
\usepackage{diagbox}
\usepackage{multirow}
\usepackage{bigdelim}
\usepackage{balance}
\usepackage{float} 

\newtheorem{lemma}{Lemma}
\newtheorem{corollary}{Corollary}
\newtheorem{example}{Example}
\theoremstyle{definition} 

\def\BibTeX{{\rm B\kern-.05em{\sc i\kern-.025em b}\kern-.08em
    T\kern-.1667em\lower.7ex\hbox{E}\kern-.125emX}}

\begin{document}

\title{Polarized Element-pair Code Based FFMA over a Gaussian 
Multiple-access Channel}
\author{Zhang-li-han Liu and Qi-yue Yu$^\dagger$, \textit{Senior Member, IEEE}
\thanks{Z.-L.-H.~Liu (email: Liuzhanglihan@stu.hit.edu.cn) and Q.-Y.~Yu (email: yuqiyue@hit.edu.cn) are affiliated with Harbin Institute of Technology, Heilongjiang, China.
}
\thanks{The work presented in this paper was supported by the National Natural Science Foundation of China under Grand No. 62071148.}
}

\maketitle
\begin{abstract}
This paper presents polarized element-pair (EP) codes for polarization-adjusted finite-field multiple-access (PA-FFMA) systems. The core innovation of FFMA systems lies in their unique processing order that exchanges the conventional sequence of channel coding and multiplexing operations, effectively solving the multiuser finite-blocklength (FBL) problem while enhancing error performance. In this architecture, EPs serve as virtual resources for user separation, where different EP codes provide distinct error performance characteristics.
The proposed polarized EP code differs from classical polar codes in one aspect that it is specifically designed for Gaussian multiple access channel (GMAC) environments rather than single-user Gaussian channels. We derive the channel capacity for this polarized EP code based FFMA system, then develop an optimal power allocation scheme to maximize multiuser channel capacity. The code construction employs the Marto Loco method for selecting the polarized index set. For decoding, we introduce two specialized algorithms. A successive cancellation list (SCL) decoder for the balanced information-parity section scenarios, and a top $L$ bifurcated minimum distance (Top$L$-BMD) decoder for small payload cases while maintaining comparable error performance. Simulations show that, for $15$ users, our system achieves a $1.25$ dB coding gain compared to the state-of-the-art polar random spreading systems.
\end{abstract}

\vspace{0.1in}

\section{Introduction}
\IEEEPARstart{F}{or} next-generation communication systems, massive machine-type communication (mMTC) plays a crucial role in accommodating a large number of low-complexity devices simultaneously \cite{MTC1,MTC2}. To support typical mMTC use cases, multiple-access (MA) schemes must be capable of handling a massive number of devices, with each device transmitting small payloads while ensuring high reliability and low transmission latency.

Current MA solutions adopt two fundamental paradigms: \textit{grant-based} and \textit{grant-free} schemes. Grant-based MA requires devices to obtain explicit authorization before transmission, trading off increased latency and control overhead for collision minimization. Conversely, grant-free MA (GFMA) allows immediate transmission without prior coordination, relying on advanced receiver-side signal processing for collision resolution \cite{5G}. This efficiency advantage has driven extensive research efforts in GFMA methodologies \cite{GrantFree}.

The implementation of GFMA critically depends on two interrelated yet independently designable aspects: first, the construction of MA coding schemes that guarantee reliable multiuser communication with controlled error performance, and second, the optimization of access protocols encompassing ALOHA variants (slotted ALOHA, $T$-fold ALOHA) along with other contention-based approaches \cite{ALOHA,s-ALOHA}. While the interaction between these components ultimately governs system performance, they admit separate design processes. This work specifically addresses the MA coding challenge, leaving protocol optimization for separate treatment.

Many studies have been conducted on multiple-access coding schemes over Gaussian multiple access channels (GMACs). Notable contributions include irregular repetition slotted ALOHA (IRSA) \cite{irsa}, BCH-based multiple access \cite{BCH1,BCH2}, interleave division multiple access (IDMA) \cite{IDMA1,IDMA2,IDMA3-POLAR}, on-off division multiple access (ODMA) \cite{ODMA1,ODMA2}, and polar-based MA schemes, among others.
In \cite{irsa}, IRSA enhances the performance of ALOHA through bipartite graph modeling and iterative interference cancellation.
In \cite{BCH1}, BCH-based multiple access employs algebraic coding principles to efficiently combine collided signals from multiple users, making it effective for handling small to medium-sized user groups. 
An IDMA system assigns each user a unique interleaver pattern. At the receiving end, it leverages turbo iterations between signal detection and decoding, enabling the recovery of multiuser bit sequences \cite{IDMA1}.
In \cite{ODMA1}, the authors propose the ODMA system, which uses sparse on-off patterns to detect users blindly, stitching messages together using tree-based rules. The structure of an ODMA system can sometimes be viewed as a special case of an IDMA system.

It is worth noting that polar-based MA schemes have gained significant popularity. Some polar-based MA systems combine polar codes with common MA schemes \cite{PS,ODMA-PC}, like spreading, ODMA, while others focus on multiuser polar codes \cite{NBPC,PC-NOMA,TUPC1,TUPC2}.
Polar codes, first given by Arikan~\cite{POLAR}, reach capacity with low complexity of coding and decoding. Later works present virous decoding algorithms to improve the error performance, such as successive cancellation list (SCL) decoding with cyclic redundancy check (CRC)~\cite{scl,ca-scl,llr-scl}, various fast successive cancellation (SC) decoding algorithms \cite{fast-sc,fast-scl,fast-scl2,fast-scl3,Sublinear1}, and etc.
Some works also study on how fast they move toward capacity~\cite{speed-rate1,speed-rate2}, better kernels~\cite{exponent,Polar-sub,pac,list-pac}, and weight distribution~\cite{Polar-Weight-Method,PC-Weight-Distribution}.
In the spreading line, the authors of \cite{PS} jointly consider polar code with random spreading. In \cite{ODMA-PC}, it combines polar code with sparse transmission pattern, so that to develop a simple and energy efficient solution. 
In the code-design line,  \cite{NBPC} explores non-binary polar codes for two-user with a simple SC decoder, and it shows that the kernel factor effects the error performance. In \cite{PC-NOMA}, the authors propose polar-coded NOMA systems which uses two-step polarization and joint decoding to outperform traditional NOMA. In \cite{TUPC1}, it joins XOR-based polar codes with physical-layer network coding so it can skip complex successive interference cancellation (SIC), and uses Monte Carlo to optimize code construction and power sharing. \cite{TUPC2} introduces polar slot codes based on T-fold irregular slotted Aloha, which are very close to the IRSA achievability bound.

Nevertheless, their error performance is fundamentally limited by the multiuser \textit{finite blocklength (FBL)} constraint \cite{FBL1,FBL2}. To overcome this inherent trade-off, we introduced \textit{finite field multiple access (FFMA)} \cite{FFMA,FFMA2,FFMA3,FFMA-ITW}, which inverts the conventional processing order of channel coding and multiplexing. This inversion facilitates the superposition of multiuser signals, resulting in a valid codeword structure.
The FFMA framework utilizes \textit{element pairs (EPs)} as virtual resources, where the Cartesian product of $J$ EPs forms an EP code. This construction supports two formulations: one is the \textit{symbol-wise} approach \cite{FFMA}, and the other is the \textit{codeword-wise} approach \cite{FFMA2}. While previous studies have predominantly focused on LDPC-based FFMA implementations, the framework is also well-suited for other linear block codes, particularly polar codes.

In this paper, we propose the \emph{polarized EP code} and, based on this, introduce \textit{polarization-adjusted FFMA (PA-FFMA)} systems. Unlike the classical polar code, which was originally designed for single-user links, the polarized EP code is specifically tailored for multiuser transmission. We present two decoding strategies: an SCL based soft decoder for large payload scenarios and a \textit{top $L$ bifurcated minimum distance} (Top$L$-BMD) hard decoder for small payload cases.
The main contributions of this paper can be summarized as follows:

\begin{enumerate}
    \item We formally define the concept of polarized EP codes and introduce the polarized EP encoder in parallel mode. The proposed systematic encoding scheme facilitates the division of output elements into distinct information and parity sections, which are then transmitted through different channel configurations.
    
    \item We conduct a capacity analysis for the PA-FFMA system, where the GMAC is decomposed into two concatenated components: a binary-input systematic channel, followed by a multiple-input nonsystematic channel. Our analysis offers a comprehensive characterization of the channel polarization process, along with the application of the Monte Carlo construction method for designing polarized EP codes. Furthermore, we introduce the power allocation scheme for the proposed PA-FFMA systems.
    
    \item For small payload scenarios, we propose a two-phase Top$L$-BMD decoding algorithm. In the first phase, a Top$L$ search is employed based on min-heap principles to process the information section and narrow the detection range. The second phase processes the parity section through re-encoding and modulation of multiuser information to identify the optimal solution.
\end{enumerate}

The rest paper is ogranized as follows. Section \ref{section2} give the definition of the polaried EP codes and its corresponding encoder in parallel mode. Section \ref{section3} presents a PA-FFMA system based on polarized EP codes. In Section \ref{construction}, we analyze channel capacity and construction
of polarized EP code. Section V shows the decoding algorithms of the proposed PA-FFMA systems. Simulation results are given in Section \ref{numeric}. Finally, some conclusions are drawn.

In this work, we employ standard mathematical notation where bold uppercase $\mathbf{X}$ denotes matrices, bold lowercase $\mathbf{x}$ represents vectors, and italic $x$ signifies scalars. The symbols $\mathbb{B}$ and $\mathbb{C}$ designate the binary field $\mathrm{GF}(2)$ and complex field respectively.

\section{Polarized EP Code over GF($2^m$)}\label{section2}
In this section, we construct \emph{polarized element-pair (EP)} code over GF($2^m$), where $m$ is a positive integer with $m \geq 2$ and GF($2^m$) is the extension ﬁeld of the prime ﬁeld GF(2). Then, 
we introduce the encoding of polarized EP code.

\subsection[short]{Kronecker Vector Spaces}
Commonly known, Polar code is constructed based on the \textit{Kronecker matrices}, whose kernel ${\bf G}^{(1)}$ is given as

\begin{equation}
{\bf G}^{(1)}=\left[ 
\begin{matrix}
    1 & 0\\
    1 & 1\\
\end{matrix}\right].
\end{equation}

Let $\kappa$ be a positive number, for $\kappa > 1$, the $\kappa$-fold
Kronecker matrix ${\bf G}^{(\kappa)}$ is the
Kronecker product of the Kenel ${\bf G}^{(1)}$ and the $(\kappa-1)$-fold Kronecker matrix 
${\bf G}^{(\kappa -1)}$, given as
\begin{equation}
    {\bf G}^{(\kappa)}={\bf G}^{(1)} \otimes {\bf G}^{(\kappa-1)} =\left[ 
        \begin{matrix}
            {\bf G}^{(\kappa-1)} & {\bf O}\\
            {\bf G}^{(\kappa-1)} & {\bf G}^{(\kappa-1)}\\
        \end{matrix}\right].
\end{equation}

The $\kappa$-fold 
Kronecker matrx $\bf G^{
(\kappa)}$ is a $2^{\kappa} \times 2^{\kappa}$
 matrix over GF(2). Each row of $\bf G^{
(\kappa)}$ is a $2^{\kappa}$-tuple over GF(2).
Hence, there are totally $2^{\kappa}$ $2^{\kappa}$-tuples in
 $\bf G^{(\kappa)}$.

For example, when $\kappa=2$, the $2$-fold Kronecker 
matrix ${\bf G}^{(2)}$ is given as 
\begin{equation}
    {\bf G}^{(2)}={\bf G}^{(1)} \otimes {\bf G}^{(1)}=
        \begin{matrix}
            _0\\_1\\_2\\_3
        \end{matrix}
            \left[\begin{matrix}
        1& 0& 0& 0\\
        1& 1& 0& 0\\
        1& 0& 1& 0\\
        1& 1& 1& 1\\
    \end{matrix}\right],
\end{equation}
which has four 4-tuples over GF(2). 
The left-hand side numbers ``0, 1, 2, 3" represent the row indices, which will be used for further discussion.

For $\kappa = 3$, the 3-fold Krnonecker matrx ${\bf G}^{(3)}$ is
\begin{equation}\label{g3}
    {\bf G}^{(3)}={\bf G}^{(2)} \otimes {\bf G}^{(1)}=
    \begin{matrix}
        _{0}\\_{1}\\_{2}\\_{3}\\_{4}\\_{5}\\_{6}\\_{7}
    \end{matrix}
    \left[\begin{matrix}
        1& 0& 0& 0& 0& 0& 0& 0\\
        1& 1& 0& 0& 0& 0& 0& 0\\
        1& 0& 1& 0& 0& 0& 0& 0\\
        1& 1& 1& 1& 0& 0& 0& 0\\
        1& 0& 0& 0& 1& 0& 0& 0\\
        1& 1& 0& 0& 1& 1& 0& 0\\
        1& 0& 1& 0& 1& 0& 1& 0\\
        1& 1& 1& 1& 1& 1& 1& 1\\
    \end{matrix}\right],
\end{equation}
which has eight 8-tuples over GF(2).

Evidently, the $\kappa$-fold Kronecker matrix is a 
full rank matrix and all its rows are linearly independent. Therefore, 
the rows of the Kronecker matrix can form a basis. All the combinations of the basis can form a 
\textit{Kronecker vector space} ${\mathbb V}_{2^{\kappa}}(2)$.

In fact, polar code is a linear block code, 
which are subspaces of the Kronecker vector spaces.
Hence, regading as the \textit{unique sum-pattern mapping 
(USPM) structural propoty} of single codeword EP (S-CWEP) 
code \cite{FFMA}, we can construct \textit{polarized EP code} 
based on the Kronecker matrix.

\subsection{Polarized EP Code}

Let $\alpha$ be a primitive element in {\rm{GF($2^m$)}}, and the powers of $\alpha$, i.e., $\alpha^0 = 1, \alpha, ..., \alpha^{2^m-2}$, give all the elements of GF($2^m$). Each element $\alpha^{l_j}$ for $0 \le l_j \le 2^m-2$ can be uniquely represented by an $m$-tuple $(a_{j,0},a_{j,1},\dots,a_{j,m-1})$ over GF($2^m$).

Let \( M \) be a positive integer, and let the \( M \) EPs be denoted by \( C_1, C_2, \dots, C_M \), where each \( C_j = ({\bf 0}, \alpha^{l_{j,1}}) \) for \( 1 \le j \le M \). The codewords \( \alpha^{l_{1,1}}, \alpha^{l_{2,1}}, \dots, \alpha^{l_{j,1}}, \dots, \alpha^{l_{M,1}} \) are selected as rows from the \( \kappa \)-fold Kronecker matrix \( {\bf G}^{(\kappa)} \), where \( m = 2^\kappa \) and \( M \le m \).
The Cartesian product of the \( M \) EPs forms an EP code \( \Psi \), defined as
\[
\Psi \triangleq C_1 \times C_2 \times \dots \times C_M = \{C_1, C_2, \dots, C_M\}.
\]

Let \( \alpha^{l_{j,1}} \) be the \( i_j \)-th row of the \( \kappa \)-fold Kronecker vector matrix \( {\bf G}^{(\kappa)} \), where \( i_j \in \mathcal{A} \).
For an arbitrary subset \( \mathcal{A} \subseteq \{0, \dots, m - 1\} \), the \textit{full-one generator matrix} \( {\bf G}_{\rm M}^{\bf 1} \) is given by
\[
{\bf G}_{\rm M}^{\bf 1} = {\bf G}_{\mathcal{A}}^{(\kappa)} = \left[ 
\alpha^{l_{1,1}}, \alpha^{l_{2,1}}, \dots, \alpha^{l_{M,1}}
\right]^{\rm T},
\]
which is an \( M \times m \) matrix representing the submatrix of \( {\bf G}^{(\kappa)} \) formed by the rows indexed by the elements of \( \mathcal{A} \).

Similarly to the polar code, 
constructing the polarized EP codes is 
to find suitable EP index set, which is 
used to approach the channel capacity of 
the FFMA system over a GMAC. The channel polarization will be discussed in the following section.

\begin{example}\label{example1}
Given a 3-fold Kronecker vector matrix ${\bf G}^{(3)}$ as shown in ({\ref{g3}}). When $M = 4$, we select the $3$-rd, $5$-th, $6$-th and $7$-th rows of ${\bf G}^{(3)}$, i.e., ${\mathcal A} = \{3, 5, 6, 7 \}$, as the $\alpha^{l_{2,1}}, \alpha^{l_{3,1}}, \alpha^{l_{4,1}}$ and $\alpha^{l_{4,1}}$. Thus, the full-one generator matrix $G_{\rm M}^{\bf 1}$ is given as
    \begin{equation}
        \begin{aligned}
        {\bf G}^{\bf 1}_{\rm M}
        ={\bf G}^{(3)}_{{\mathcal A}}
        =\left[\begin{matrix}
            \alpha^{l_{1,1}}\\ \alpha^{l_{2,1}}\\ \alpha^{l_{3,1}}\\         
            \alpha^{l_{4,1}}\\
        \end{matrix}\right]
        = \left[\begin{matrix}
            1& 1& 1& 1& 0& 0& 0& 0\\
            1& 1& 0& 0& 1& 1& 0& 0\\
            1& 0& 1& 0& 1& 0& 1& 0\\
            1& 1& 1& 1& 1& 1& 1& 1\\
        \end{matrix}\right],
        \end{aligned}
    \end{equation}
which is a $4 \times 8$ matrix. The corresponding 4-user EPs are given as:
    \begin{equation}
        \begin{aligned}
            C_1 &= ({\bf 0}, \alpha^{l_{1,1}}) ,\quad
            C_2 = ({\bf 0}, \alpha^{l_{2,1}}) ,\\
            C_3 &= ({\bf 0}, \alpha^{l_{3,1}}) ,\quad
            C_4 = ({\bf 0}, \alpha^{l_{4,1}}),
        \end{aligned}
    \end{equation}
    which can totally support \(2^4=16\) EP codewords.      
    $\blacktriangle \blacktriangle$
\end{example} 


\subsection{Encoding of Polarized EP Code}
Suppose the number of users is $J$, and each user transmitting $K$ bits, with the assumption $J \cdot K \leq M$. In this paper, the EP encoder operates in parallel mode. It indicates that, $K$ different EPs, i.e., $C_{(j - 1) \cdot K + 1}, C_{(j - 1) \cdot K + 2}, ..., C_{j \cdot K}$, are assigned to the $j$-th user. Therefore, the EP encoder map each bit ${b}_{j,k}$ of the $j$-th user into an element ${u}_{j,k}$ as follow:
\begin{equation}
    \begin{aligned}        
        {u}_{j,k}&={\text F_{\text B2q}}(b_{j,k})=b_{j,k} \odot  C_{(j - 1) \cdot K + k + 1}
        \\&=
        \begin{cases}
            {\bf 0}, &\text{ if } b_{j,k}=0 \\
            \alpha^{l_{(j-1)\cdot K+k + 1,1}}, &\text{ if } b_{j,k}=1
        \end{cases},
    \end{aligned}
\end{equation}
where $b_{j,k} \odot  C_{(j - 1) \cdot K + k + 1}$ is a switch function and ${u}_{j,k}$ is the output element. 
After the ${\rm F}_{{\rm B}2q}$ mapping, the $K$ elements $u_{j, 0}, u_{j, 1}, ..., u_{j, K-1}$ are summed into an $m$-tuple, denoted as ${\bf c}_j=(c_{j,0},\dots,c_{j,m-1})$, given as
\begin{equation}
    {\bf c}_j = \bigoplus_{k=0}^{K-1}{{u}_{j,k}},\\
\end{equation}
which is referred to as a \textit{finite-field sum pattern (FFSP) block}.
The sum of $J$ users' output elements is also an FFSP block, defined as \({\bf v} = (v_0,..., v_{m-1})\), i.e.,
\begin{equation}
        \label{FFSP block}
        {\bf v} = \bigoplus_{j=1}^{J}{\mathbf c_j}=\bigoplus_{j=1}^{J}\bigoplus_{k=0}^{K-1}{u_{j,k}}
        \overset{(a)}{=} {\bf w} \cdot {\bf G}_{\rm M}^{\bf 1}
        \overset{(b)}{=} {\bf d} \cdot {\bf G}^{(\kappa)},
\end{equation}
where $\bf w$ is \textit{user information block} with length $JK$. It has two equivalent representations: the block-level form $\mathbf{w} = (\mathbf{b}_1, \ldots, \mathbf{b}_J)$ where each $\mathbf{b}_j$ denotes a $K$-length information block of the $j$-th user, and the symbol-level form $\mathbf{w} = (w_0, \ldots, w_{JK-1})$. Besides, let a $1 \times m$ vector be denoted as $\mathbf{d} = (d_0, \ldots, d_{m-1})$, where active positions $i \in \mathcal{A}$ carry information bits from $\mathbf{w}$ such that $d_i = w_k$ for some $k \in \{0,\ldots,JK-1\}$, while inactive positions $i \notin \mathcal{A}$ are frozen to $0$. This construction ensures $\mathbf{w}$ is embedded as a proper sub-vector of $\mathbf{d}$, maintaining the original information content while enabling length extension through zero-padding. The index set $\mathcal{A}$ precisely determines the mapping between $\mathbf{w}$ and the non-zero elements of $\mathbf{d}$.

However, for the elements of certain users, it is possible that certain bits are always zero, as shown in Example 2.
\begin{example}\label{example2}    
Suppose there is one user, with transmitting $K = 2$ bits. The generator matrix has been given in Example 1. We explore the EP encoder in parallel mode, and assign two EPs, e.g., $C_1$ and $C_2$ to the user. Next, we can calculate the output element which is the sum of the two elements according to Eq. (\ref{FFSP block}). Consider ${\bf b}_1 \in \{00, 01, 10, 11\}$, we have
    \[
        \mathbf{c}_1 =
        \begin{cases}
        \mathbf{0}  &= \quad (0\ 0\ 0\ 0\ 0\ 0\ \textcolor{red}{0}\ \textcolor{red}{0}),  \quad\text{if } \mathbf{b}_1 = 00, \\[1ex]
        \alpha^{l_{1,1}}  &= \quad (1\ 1\ 1\ 1\ 0\ 0\ \textcolor{red}{0}\ \textcolor{red}{0}),  \quad\text{if } \mathbf{b}_1 = 01, \\[1ex]
        \alpha^{l_{2,1}}  &=\quad (1\ 1\ 0\ 0\ 1\ 1\ \textcolor{red}{0}\ \textcolor{red}{0}),  \quad\text{if } \mathbf{b}_1 = 10, \\[1ex]
        \alpha^{l_{1,1}} \oplus \alpha^{l_{2,1}}  &= \quad (0\ 0\ 1\ 1\ 1\ 1\ \textcolor{red}{0}\ \textcolor{red}{0}),  \quad\text{if } \mathbf{b}_1 = 11.
        \end{cases}
    \]

    In this case, the four possible combinations of the two bits result in elements where the last two bits are always zero as indicated by the red 0 in the equation above. This is because the last two columns of \(\alpha^{l_{1,1}}\) and \(\alpha^{l_{2,1}}\) are both zero.
    $\blacktriangle \blacktriangle$
\end{example}
From the above example, we observe that the last two bits of the elements are always zero. 
Consequently, these non-information bits can be eliminated to optimize transmission power efficiency. However, in a non-systematic encoding scheme, the specific bits to be pruned are dynamically determined by the assigned EPs. This dependency introduces significant implementation complexity, as the omission positions vary according to the instantaneous coding configuration.

\subsection{Systematic Polarized EP codes}
Hence, in this paper, we transform the matrix \( {\bf G}_{\rm M}^1 \) into its systematic form to address this issue.
Let the full-one generator matrix in systematic form ${\bf G}_{\rm M, sym}^{\bf 1}$ be denoted as
\begin{equation}   
    \mathbf{G}_{\mathrm{M,sym}}^{\mathbf{1}} = \left[ 
        \begin{array}{c:c}
            \mathbf{I} & \mathbf{Q} 
        \end{array}
     \right],
\end{equation}
where \(\bf Q\) is the parity section of the generator matrix, which is an \(M \times R\) matrix where \(R = m - M\). \(\bf I\) represents the information section of the generator matrix, whose size is \(M \times M\).
Refer to \cite{System-PC}, we can transform a non-systematic generator matrix \({\bf G}_{\rm M}^{\bf 1}\) into a systematic generator matrix \({\bf G}_{\rm M, sym}^{\bf 1}\) by the following two steps.

\begin{itemize}
    \item 
    \textbf{Step 1:} Left-multiply the matrix  
    \(
        \left(\mathbf{G}_{\mathcal{A}\mathcal{A}}^{(\kappa)}\right)^{-1},
\)  
yielding a new matrix:  
\(
    \left(\mathbf{G}_{\mathcal{A}\mathcal{A}}^{(\kappa)}\right)^{-1} \mathbf{G}_{\mathrm{M}}^{\mathbf{1}}.
\) 
    Here, $\mathbf{G}_{\mathcal{A}\mathcal{A}}^{(\kappa)}$ denotes the $M\times M$ submatrix of $\mathbf{G}^{(\kappa)}$ formed by the rows and columns indexed by $\mathcal{A}$.

    \item \textbf{Step 2:} By column operations that group the information bits first (and the remaining bits as parity), the full-one generator matrix $\mathbf{G}_{\mathrm{M}}^{1}$ is converted into its systematic form $\mathbf{G}_{\mathrm{M},\mathrm{sym}}^{1}$.

\end{itemize}

\begin{example}\label{example3}
Consider the systematic form generator matrix \( {\bf G}_{\rm M, sym}^{\bf 1} \) derived from the non-systematic generator matrix \( {\bf G}_{\rm M}^{\bf 1} \) in Example \ref{example1}. By left-multiplying \( {\bf G}_{\rm M}^{\bf 1} \) with \( \left( {\bf G}_{{\mathcal A}{\mathcal A}}^{(\kappa)} \right)^{-1} \), we obtain the following result:
    \begin{equation}
        \begin{aligned}
            ({{\bf G}_{{\mathcal A}{\mathcal A}}^{(\kappa)}})^{-1} {\bf G}_{\rm M}^{\bf 1} &= 
            \begin{bmatrix}
                1 & 0 & 0 & 0 \\
                0 & 1 & 0 & 0 \\
                0 & 0 & 1 & 0 \\
                1 & 1 & 1 & 1
            \end{bmatrix}
            \begin{bmatrix}
                1 & 1 & 1 & 1 & 0 & 0 & 0 & 0 \\
                1 & 1 & 0 & 0 & 1 & 1 & 0 & 0 \\
                1 & 0 & 1 & 0 & 1 & 0 & 1 & 0 \\
                1 & 1 & 1 & 1 & 1 & 1 & 1 & 1
            \end{bmatrix}
            \\&= 
            \left[\!\!\!
            \begin{array}{cccccccc}
                1 & 1 & 1 & 1 & 0 & 0 & 0 & 0 \\
                1 & 1 & 0 & 0 & 1 & 1 & 0 & 0 \\
                1 & 0 & 1 & 0 & 1 & 0 & 1 & 0 \\
                0 & 1 & 1 & 0 & 1 & 0 & 0 & 1 \\
            \end{array}
            \!\!\!
            \right].  
            \label{GG}                
        \end{aligned}
    \end{equation}
    Next, the columns of the \( ({{\bf G}_{{\mathcal A}{\mathcal A}}^{(\kappa)}})^{-1} {\bf G}_{\rm M}^{\bf 1}\) with column indices in \(\mathcal{A}\) form an identity matrix. We only need to assemble the columns in \(\mathcal{A}\) to obtain the identity matrix. For the matrix in Eq. (\ref{GG}), we swap the $4$-{th} and $5$-{th} columns as follows:
    \begin{equation}
        \begin{aligned}            
            {\bf G}_{\rm M,sym}^{\bf 1} &= 
            {\bf I}_8' ({{\bf G}_{{\mathcal A}{\mathcal A}}^{(\kappa)}})^{-1} {\bf G}_{\rm M}^{\bf 1} \\&=\left[\begin{matrix}
                \alpha^{l_{1,1}}\\ \alpha^{l_{2,1}}\\ \alpha^{l_{3,1}}\\ \alpha^{l_{4,1}}\\
            \end{matrix}\right]= 
            \left[\!\!\!
            \begin{array}{cccc:cccc}
                1 & 0 & 0 & 0 & 1 & 1 & 1 & 0  \\
                0 & 1 & 0 & 0 & 1 & 1 & 0 & 1  \\
                0 & 0 & 1 & 0 & 1 & 0 & 1 & 1  \\
                0 & 0 & 0 & 1 & 0 & 1 & 1 & 1  \\
            \end{array}
            \!\!\!
            \right].
            \label{IGG}
        \end{aligned}
    \end{equation}
    Here, $\mathbf{I}_8'$ is the identity matrix $\mathbf{I}_8$ with its columns permuted such that columns 3, 5, 6, and 7 are moved to positions 0, 1, 2, and 3, respectively.
    \(\blacktriangle \blacktriangle\)
\end{example}

\begin{figure}[t]
    \centering
    \includegraphics[width=0.8\linewidth]{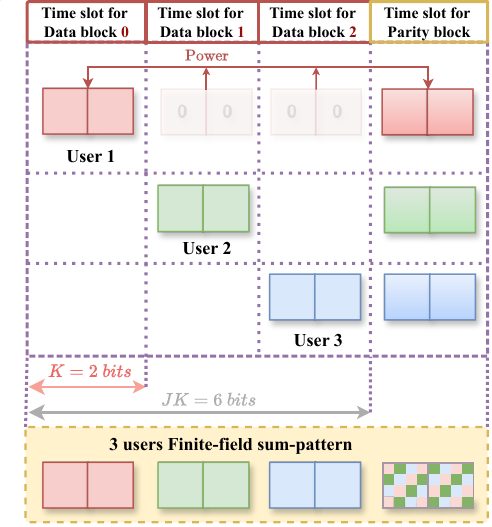}
    \caption{A diagram of the systematic FFMA system, where \( J = 3 \), \( K = 2 \), \( m = 2^3 \).}
    \label{fig:system_FFMA}
\end{figure}

\subsection{CRC-Aided Systematic Polarized EP codes}

In \cite{ca-scl}, CRC-aided polar codes have been extensively studied, demonstrating exceptional performance with successive cancellation list decoding. Building on these results, we extend the systematic polarized EP code to its augmented form in this paper.

For the CRC-aided systematic polarized EP code, the construction involves two sequential steps: first, forming an $M \times (M + C_L)$ systematic CRC encoding matrix, and then establishing a $(M + C_L) \times m$ generator matrix $\mathbf{G}_{\text{crc}}$. 
This process can be summarized in the following corollary.
\begin{corollary}
Since $\mathbf{G}^{1}_{\rm{M,sym}}$ is a full-one generator matrix, it has full row rank. If the CRC matrix $\mathbf{G}_{\text{crc}}$ is also full rank, then the multiplexing of the two matrices, i.e.,  
    \begin{equation}\label{G_syspolar}
        \mathbf{G}^{1}_{\rm{M,sym,crc}} = \mathbf{G}_{\rm{crc}} \cdot \mathbf{G}^{1}_{\rm{M,sym}},
    \end{equation}
results in a full row rank matrix of size $M \times m$. Therefore, $\mathbf{G}^{1}_{\rm{M,sym,crc}}$ is the full-one generator matrix of the CRC-aided systematic polarized EP code, and the corresponding EP code is defined by $\Psi_{\rm pl,crc}$.
\end{corollary}
    
Hence, the CRC-aided systematic polarized EP code, \( \Psi_{\rm pl,crc} \), can still leverage the previously described EP encoder. Unlike the non-systematic polarized EP code, the output element \( \mathbf{c}_j \) of the systematic polarized EP code \( \Psi_{\rm pl,crc} \) is divided into two distinct sections: the information section \( \mathbf{c}_{j, \rm inf} \) with length \( M \) and the parity section \( \mathbf{c}_{j, \rm red} \) with length \( R = m - M \), i.e., \( \mathbf{c}_j = (\mathbf{c}_{j, \rm inf}, \mathbf{c}_{j, \rm red}) \).

The information section \( \mathbf{c}_{j, \rm inf} \) of \( \mathbf{c}_j \) for the \( j \)-th user consists of an \( M \)-tuple, which is further divided into \( J \) blocks, each containing \( K \) bits, such that \( M = K \times J \). For the \( j \)-th user, the \( j \)-th block is used to transmit the bit sequence \( \mathbf{b}_j \), while the other blocks are set to zero, as shown by \( \mathbf{c}_{j, \rm inf} = (\mathbf{0}, \dots, \mathbf{0}, \mathbf{b}_j, \mathbf{0}, \dots, \mathbf{0}) \). As a result, these fixed zero bits can be eliminated, allowing for the allocation of more power to the useful information section, further enhancing the process of \textit{polarization}.

In the parity section, if \( K \) is small, some parity bits may always be zero. In such cases, omitting these zero parity bits could be beneficial for reducing power consumption. However, as the locations of the parity bits depend on the selection of the index set \( \mathcal{A} \), this paper does not explore the possibility of further reducing the power allocated to the parity section.

\section{System Model}\label{section3}

\begin{figure*}[!t]
    \centering
    \includegraphics[width=0.87\textwidth]{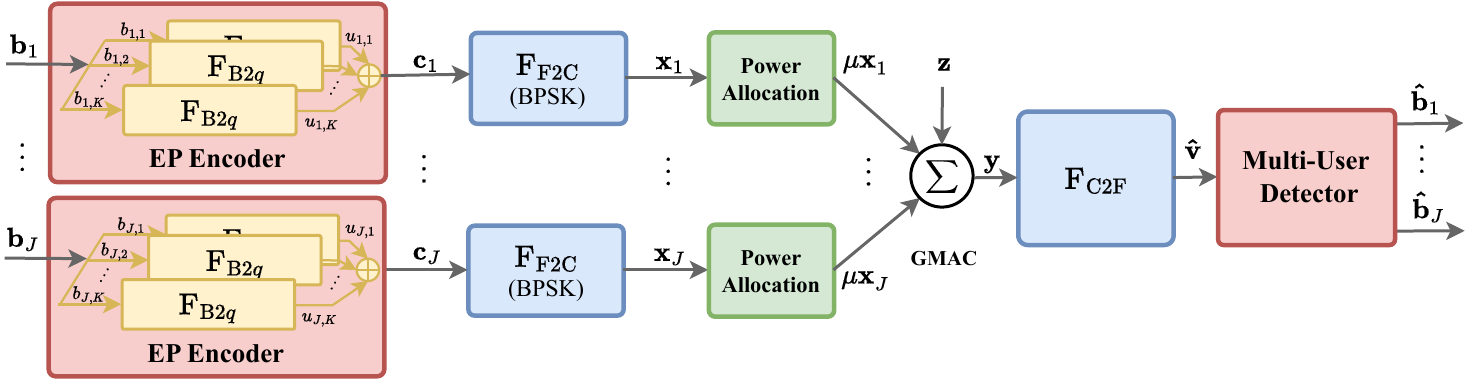}
    \caption{An FFMA system over a GMAC, with the transmitter comprising an EP encoder, a transform function ${\rm F_{F2C}}$, and a PA module, and the receiver is consisted of a transform function ${\rm F_{C2F}}$ and a multi-user detector.}
    \label{fig:FFMA_system}
\end{figure*}

This section introduces a polarized EP code based FFMA system over a GMAC, where the transmitter includes an EP encoder, a transform function ${\rm F_{F2C}}$, and an (operation) power allocation (PA) module. At the receiving end, it is consisted of a transform function ${\rm F_{C2F}}$ and a multi-user detector (MUD), as shown in Fig. \ref*{fig:FFMA_system}.

\subsection{Transmitter of an FFMA system}

Let ${\bf b}_j=(b_{j,1},b_{j,2},\dots,b_{j,k},\dots,b_{j,K})$ be the bit-sequence of the $j$-th user, where $1 \le j \le J$, $0\le k \le K - 1$. 
For the $j$-th user, the bit sequence \(b_j\) is encoded into a output element 
${\bf c}_j=({c}_{j,1},\dots,{c}_{j,i},\dots,{c}_{j,m})$, which is an \(m\)-tuple over GF($2$).
Then, the output element ${\bf c}_j$ of the $j$-th user is transformed to a complex-field vector ${\bf x}_j \in \mathbb{C}^{1 \times m}$ through a mapping function ${\rm F}_{\mathrm{F2C}}$, denoted as ${\bf x}_j = {\rm F}_{\mathrm{F2C}}({\bf c}_j)$. The transmit signal in symbol-level form is  
\(
        {\bf x}_j = (x_{j,0},\dots, x_{j,i}, \dots,x_{j,m-1}),
\)
with the $i$-th component 
\begin{equation}
    x_{j,i} = 
    \begin{cases}
        0, & \text{if } c_{j,i} \text{ is permanently zero} \\
        1 - 2c_{j,i}, & \text{otherwise}
    \end{cases}.
\end{equation} 
Next, the transmit signal x is operated by power allocation. Define \textit{polarization adjusted vector} in symbol form as:
\[
\boldsymbol{\mu}=(\mu_0,\dots,\mu_{m-1})=(\mathbf{0}, \underbrace{\mu_{\mathrm{inf}},\dots,\mu_{\mathrm{inf}}}_{K}, \mathbf{0},\;\underbrace{\mu_{\mathrm{red}},\dots,\mu_{\mathrm{red}}}_{R})
\]
where \(\mu_{\rm inf}\) and \(\mu_{\rm red}\) are polarization adjusted factors factors for the information section and the parity section, respectively.

Let \( P_{\mathrm{avg}} \) denote the average transmit power per symbol. To ensure the total transmit power remains constant at \( mP_{\mathrm{avg}} \), as in a polar-coded system, the power constraint is given by:
\begin{equation} \label{e.PowerConstraint}
mP_{\mathrm{avg}} = K \cdot \mu_{\mathrm{inf}} P_{\mathrm{avg}} + R \cdot \mu_{\mathrm{red}} P_{\mathrm{avg}}.
\end{equation}
Thus, the \textit{polarization transmit signal} \( {\bf x}_{j, \rm PA} \) is defined as:
\[
{\bf x}_{j, \rm PA} = \sqrt{P_{\mathrm{avg}} \boldsymbol{\mu}} \circ {\bf x}_{j},
\]
where \( \circ \) denotes the Hadamard product, and the signal is then transmitted to the GMAC.

\subsection{Receiver of an FFMA system}
At the receiver, the superimposed symbol vector is denoted as \( \mathbf{y} = (y_0, y_1, \dots, y_{m-1}) \), where the elements satisfy the following equation:
\begin{equation}\label{y_i}
y_i = \sqrt{P_{\mathrm{avg}} \mu_i} \cdot \sum_{j=1}^J x_{j,i} + z_i = \sqrt{P_{\mathrm{avg}} \mu_i} \cdot r_i + z_i.
\end{equation}
Here, \( z_i \) represents the additive white Gaussian noise (AWGN), which is modeled as \( z_i \sim \mathcal{N}(0, N_0 / 2) \), where \( N_0 \) is the noise power spectral density. The term \( r_i \) represents the superimposed signal from the \( J \)-users, given by:
\begin{equation}\label{r_i}
r_i = \sum_{j=1}^{J} x_{j,i}.
\end{equation}
This defines the \( i \)-th component of the \emph{complex-field sum-pattern (CFSP)} block, denoted as \( \mathbf{r} = (r_0, r_1, \dots, r_{i}, \dots, r_{m-1}) \).

According to \cite{FFMA}, the CFSP symbol \( r_i \) of the \( J \)-user belongs to the set:
\[
\Omega_r \triangleq \{-J, -J+2, \dots, J-2, J\},
\]
with the corresponding probabilities given by:
\[
\mathcal{P}_r \triangleq \left\{ \frac{C_{J}^{0}}{2^J}, \frac{C_{J}^{1}}{2^J}, \dots, \frac{C_{J}^{J}}{2^J} \right\}.
\]
Next, the received CFSP signal is transformed into its corresponding FFSP symbol via the \textit{complex to finite-field (C2F)} function, denoted as \( v_i = {\text{F}}_{\text{C2F}}(r_i) \). The one-to-one mapping set for \( \Omega_r \) is given by:
\(
\Omega_v = \{1, 0, 1, 0, \dots \},
\)
where $0$ and $1$ alternate, as described in \cite{FFMA}.

Then, the posterior probability of $v_i = 0$ and $v_i = 1$ are given as
\begin{equation}
    \label{PZ}
    \begin{aligned}
        &P(v_i=0|y_i) \\ &= \sum_{\substack{\nu=0 \\ \nu \leftarrow \nu+2}}^{\nu \le J}{\frac{P_r(\nu)}{p(y_i)\sqrt{\pi N_0}}\exp\left(-\frac{[y_i-\sqrt{P_{\rm avg} \mu_i} \Omega_r(\nu)]^2}{N_0}\right)}
    \end{aligned},
\end{equation}
and
\begin{equation}
    \label{PO}
    \begin{aligned}
        &P(v_i=1|y_i) \\ &= \sum_{\substack{\nu=1 \\ \nu \leftarrow \nu+2}}^{\nu \le J}{\frac{P_r(\nu)}{p(y_i)\sqrt{\pi N_0}}\exp\left(-\frac{[y_i-\sqrt{P_{\rm avg} \mu_i}\Omega_r(\nu)]^2}{N_0}\right)}.        
    \end{aligned}
\end{equation}
While, \(p(y_i)\) is given by
\begin{equation}\label{py}
    p(y_i)=\sum_{\nu=0}^{J}\frac{{\mathcal P}_r(\nu)}{\sqrt{\pi N_0}}\exp\left(-\frac{[y_i-\sqrt{P_{\rm avg} \mu_i}\Omega_r(j)]^2}{N_0}\right).
\end{equation}

The obtained posterior probabilities are used to calculate the initial log-likelihood ratio (LLR) for subsequent decoding. The initial LLR is given by:
\begin{equation}\label{LLR}
    \begin{aligned}
        &l(y_i)=\ln \frac{P(v_i=0|y_i)}{P(v_i=1|y_i)}=\ln \frac{P(y_i|v_i=0)}{P(y_i|v_i=1)}\\
        &=\ln \frac{\sum_{\nu=0,\nu+2}^{\nu \le J}{P_r(\nu)\exp\left(-\frac{[y_i-\sqrt{P_{\rm avg} \cdot \mu_i} \cdot\Omega_r(\nu)]^2}{N_0}\right)}}{\sum_{\nu=1,\nu+2}^{\nu \le J}{P_r(\nu)\exp\left(-\frac{[y_i-\sqrt{P_{\rm avg} \cdot \mu_i} \cdot\Omega_r(\nu)]^2}{N_0}\right)}}
    \end{aligned}.
\end{equation}
The resulting LLRs are then used by the MUD for decoding, which will be discussed in further detail in Section \ref{decode}.

\section{Channel Capacity and Construction of Polarized EP code}\label{construction}
In this section, we analyze channel capacity and construction of polarized EP code.

\subsection{Channel Capacity of a GMAC}
In the framework of the systematic FFMA system, it is composed of two sections: the information section, which passes through an AWGNC, and the parity section, which represents the superposition of signals from $J$ users. As a result, the equivalent channel for the parity section is a GMAC. Notably, the AWGNC can be considered a special case of the GMAC when $J = 1$. Therefore, we utilize the GMAC to analyze the channel capacity of the FFMA system.

\begin{figure}[t]
    \centering
	\includegraphics[width=0.8\linewidth]{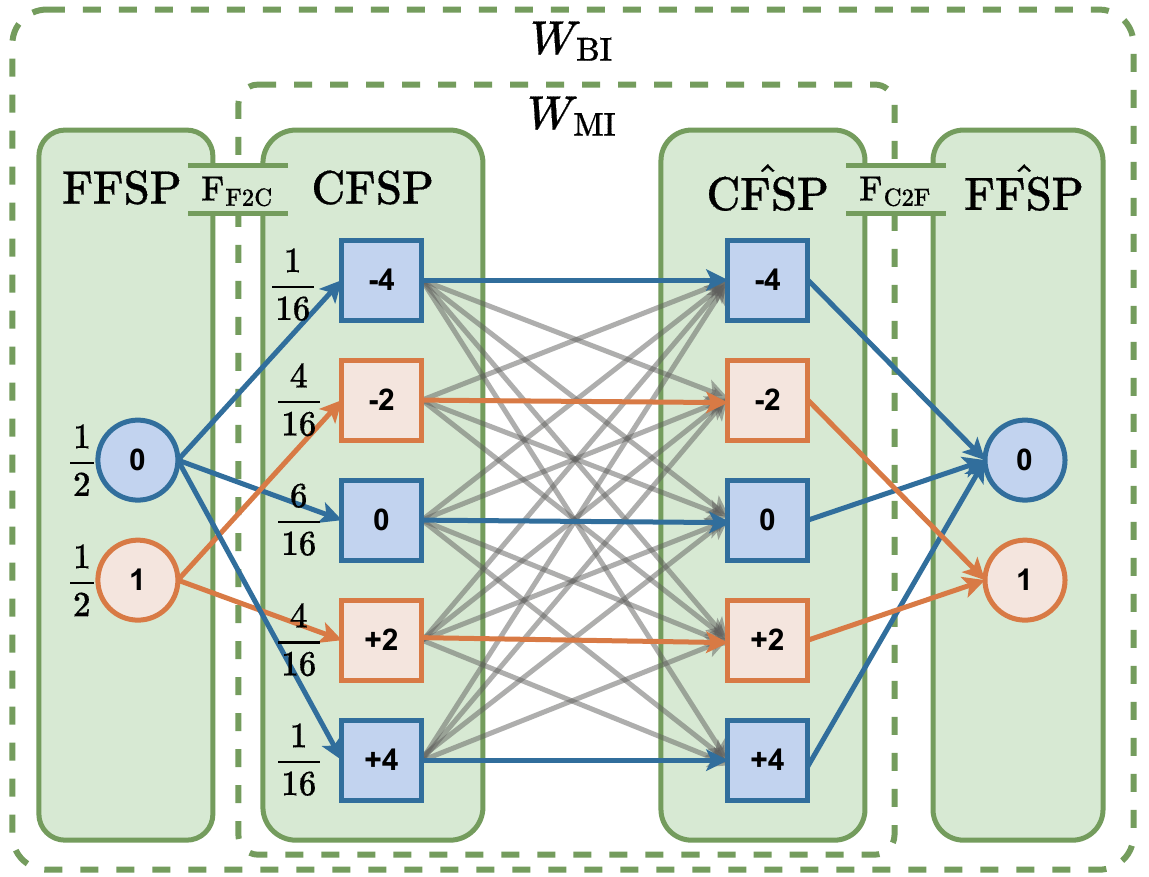}
	\caption{The GMAC can be modeled as a cascade of MI-NSC and MI-ASC, where $J = 4$.}
	\label{fig:W_MB}
\end{figure}

In fact, for the proposed FFMA system, the GMAC can be modeled as a cascade of direct channels. The first is a \textit{binary-input approximate-symmetric channel (BI-ASC)}, followed by a \textit{multiple-input non-symmetric channel (MI-NSC)}. The direct inputs of the BI-ASC come from a binary field with 2 possible inputs, while the direct inputs of the MI-NSC come from a complex field with $J + 1$ possible inputs, as shown in Fig. \ref{fig:W_MB}.

\subsubsection{Capactiy of the BI-ASC}
When the input symbols $0$ and $1$ of the BI-ASC have equal probabilities, the output probabilities of $0$ and $1$ are equal if the number of users is even. However, if the number of users is odd, the output probabilities of $0$ and $1$ are close but not completely equal \cite{FFMA3}. Therefore, we assume that the input symbols $0$ and $1$ are uniformly distributed, each with a probability of $0.5$.

We use $W_{\text{BI}}: \mathcal{V} \to \mathcal{L}$ to represent a BI-ASC, where $\mathcal{V}$ is the input alphabet, and $\mathcal{L}$ is the output space. The input symbols $v_i \in \mathcal{V}$ are defined in Eq. (\ref{FFSP block}), and the output symbols $l_i \in \mathcal{L}$ are for $M \leq i \leq m - 1$. Here, $l_i$ represents the LLR of the estimated symbol $\hat{v}_i$. 
The BI-ASC channel is approximated as a BSC channel, with its capacity given by
\begin{equation}
    C_{\rm BI} = \arg\max_{p(y_i)} {\rm I}(v_i; \hat{v}_i) \approx 1 - {\rm H}(p_e),
\end{equation}
where ${\rm H}(p_e)$ is the binary entropy function:
\begin{equation}
    {\rm H}(p_e) = -p_e \log_2(p_e) - (1 - p_e) \log_2(1 - p_e),
\end{equation}
and the symbol error rate $p_e$ is given as
\begin{equation}
    \begin{aligned}
    p_e &\approx 2 \sum_{j=0}^{J-1} \mathcal{P}_r(j) Q \left( \sqrt{\gamma} + \frac{1}{2\sqrt{\gamma}} \ln \left( \frac{J + j + 1}{J - j} \right) \right) \\
    & \quad + 2 \sum_{j=0}^{J-1} \mathcal{P}_r(j+1) Q \left( \sqrt{\gamma} - \frac{1}{2\sqrt{\gamma}} \ln \left( \frac{J + j + 1}{J - j} \right) \right),
    \end{aligned}
\end{equation}
where the SNR is defined as $\gamma = \frac{\mu_{\rm red} P_{\rm avg}}{\sigma^2}$, where $\sigma^2 = N_0/2$.

The bit error rate (BER) performance of the proposed FFMA system is primarily determined by the adjacent symbols. As the number of users increases, the received average power increases, but the distance between adjacent symbols remains unchanged, resulting in a relatively stable BER.

\subsubsection{Capactiy of the MI-NSC}
We define an MI-NSC as \( W_{\text{MI}}: \mathcal{R} \to \mathcal{Y} \), where \( \mathcal{R} \) represents the input alphabet, \( \mathcal{Y} \) is the output alphabet, and \( r_i \in \mathcal{R} \) is defined in Eq. (\ref{r_i}), while \( y_i \in \mathcal{Y} \) is defined in Eq. (\ref{y_i}). The capacity of the MI-NSC is then given by:
\begin{equation}
    \begin{aligned}
    C_{\text{MI}} &= \arg\max_{p(y_i)} {\rm I}(r_i; y_i) 
    = {\rm H}(y_i) - {\rm H}(y_i | r_i) \\
    &= \int_{-\infty}^{+\infty} p(y_i) \log_2 \frac{1}{p(y_i)} \, dy_i - {\rm H}(z),
\end{aligned}
\end{equation}
where \( p(y_i) \) is given in Eq. (\ref{py}).
When the effect of noise is not considered, the maximum capacity is given by:
\begin{equation} \label{e.C_MI}
    C_{\text{max, MI}} = {\rm H}({\mathcal R}) \leq \sum_{\nu = 0}^{J} {\mathcal P}_r(\nu) \log \frac{1}{{\mathcal P}_r(\nu)},
\end{equation}
where \( {\rm H}(\mathcal R) \) denotes the entropy of the CFSP variable with \( {\mathcal R} \in \Omega_r \), and \( {\mathcal P}_r(\nu) \) represents the probability of each element in \( \Omega_r \).

From (\ref{e.C_MI}), it can be observed that the channel capacity increases with the number of users. In a special case, as the number of users approaches infinity, i.e., \( J \to \infty \), the distribution of the input signal approaches a Gaussian distribution, since \( r_i \) is the sum of \( J \) uniformly distributed variables \( x_{j,i} \), where \( x_{j,i} \in \{-1,+1\} \) follows a Bernoulli distribution. According to the De Moivre-Laplace theorem, as \( J \to \infty \), \( r_i \) is approximately \( N(0, JP) \). As a result, the channel capacity approaches the Shannon capacity of \( J \) users, given by
\begin{equation}
C_{\text{MI}} = \frac{1}{2} \log_2 \left( 1 + \frac{J \mu_{\text{red}} P_{\text{avg}}}{\sigma^2} \right),
\end{equation}
which is in accordance with multiuser information theory \cite{Thomas}.

\subsubsection{Channel capacity of a CRC-aided Systematic Polarized EP code based FFMA system}
For a systematic FFMA system, the channel corresponding to the information section follows an AWGNC model, while the channel corresponding to the parity section is MI-NSC. Therefore, the total capacity of the system is given by \cite{FFMA3}
\begin{equation}
    C_{\mathrm{tot}} = \frac{JK}{2} \log_2 \left( 1 + \frac{\mu_{\rm inf} P_{avg}}{\sigma^2} \right) 
    + \frac{R}{2} \log_2 \left( 1 + \frac{J \mu_{\rm red} P_{avg}}{\sigma^2} \right).
\end{equation}
The polarization-adjusted factors, \( \mu_{\rm inf} \) and \( \mu_{\rm red} \), must satisfy the power allocation constraint given in (\ref{e.PowerConstraint}). For convenience in discussion, we define the \textit{polarization-adjusted scaling factor (PAS)} as the ratio of \( \mu_{\rm inf} \) to \( \mu_{\rm red} \), i.e., $\mu_{\rm pas} = \frac{\mu_{\rm inf}}{\mu_{\rm red}}$.
In Section \ref{numeric}, we will use Monte Carlo simulations to determine the optimal PAS, \( \mu_{\rm pas} \), that maximizes the channel capacity.

\subsection{Construction of Polarized Element-pair Codes}
This section explains the construction of the polarized EP code, determining the EP index set 
$\mathcal{A}$. The capacity of each bit channel 
is calculated, and the indices of the best channels are selected to form the elements of $\mathcal{A}$. 

The Gaussian approximation method in \cite{GA}, designed for single-user scenarios, proves inadequate for multiuser FFMA systems due to three key factors:
(1) The parity section constitutes an asymmetric channel where all-zero codewords are invalid \cite{asymmetric};
(2) Non-identical channel distributions caused by distinct capacities between information bits and parity bits;
(3) Computational infeasibility of density evolution analysis due to its complexity.

To address this, we employ Monte Carlo simulations to calculate polarized channel capacities. Under the perfect-assumption SC decoding framework, BER performance is evaluated with different polarization adjusted factors allocated to information and parity sections \cite{FFMA2}.

We perform Monte Carlo simulations by feeding LLR values $l_i$ from the $W_{\mathrm{BI}}$ channel output into an ideal SC decoder, generating output LLRs $l_m^{(i)}$ to statistically approximate their probability distributions. Here, $l_m^{(i)}$ denotes the LLR of the $i$-th bit channel in the $m$ polarized bit channels, timely the LLR of the \(d_i\). Based on the polarized channel capacity calculation formula from \cite{howtoconstruct}:
\begin{equation}
    I(W_m^{(i)}) = \int_{0}^{\infty} p(l_m^{(i)}) C_{\text{BSC}}(l_m^{(i)}) \, dl_m^{(i)},
\end{equation}
where $C_{\text{BSC}}$ denotes the equivalent BSC channel capacity and is defined as:
\begin{equation}
    \begin{aligned}
        C_{\text{BSC}}(l_m^{(i)}) &= 1 - \frac{\lambda(l_m^{(i)})}{\lambda(l_m^{(i)}) + 1} \log_2\left(1 + \frac{1}{\lambda(l_m^{(i)})}\right) \\
        &\quad - \frac{1}{\lambda(l_m^{(i)}) + 1} \log_2\left(1 + \lambda(l_m^{(i)})\right),
    \end{aligned}
\end{equation}
with the LLR $\lambda(l_m^{(i)}) \triangleq \frac{p(l_m^{(i)}|d_i=0)}{p(l_m^{(i)}|d_i=1)}$, where $d_i$ is defined by the equation labeled $(b)$ in Eq. (\ref{FFSP block}).
Clearly, the PDF of the LLR output from the polarized channel directly impacts its channel capacity.

The following example illustrates different numbers of users lead to distinct initial LLR pdfs, which subsequently result in varying output LLR pdfs.

\begin{figure}[t]
    \centering
    \includegraphics[width=0.9\linewidth]{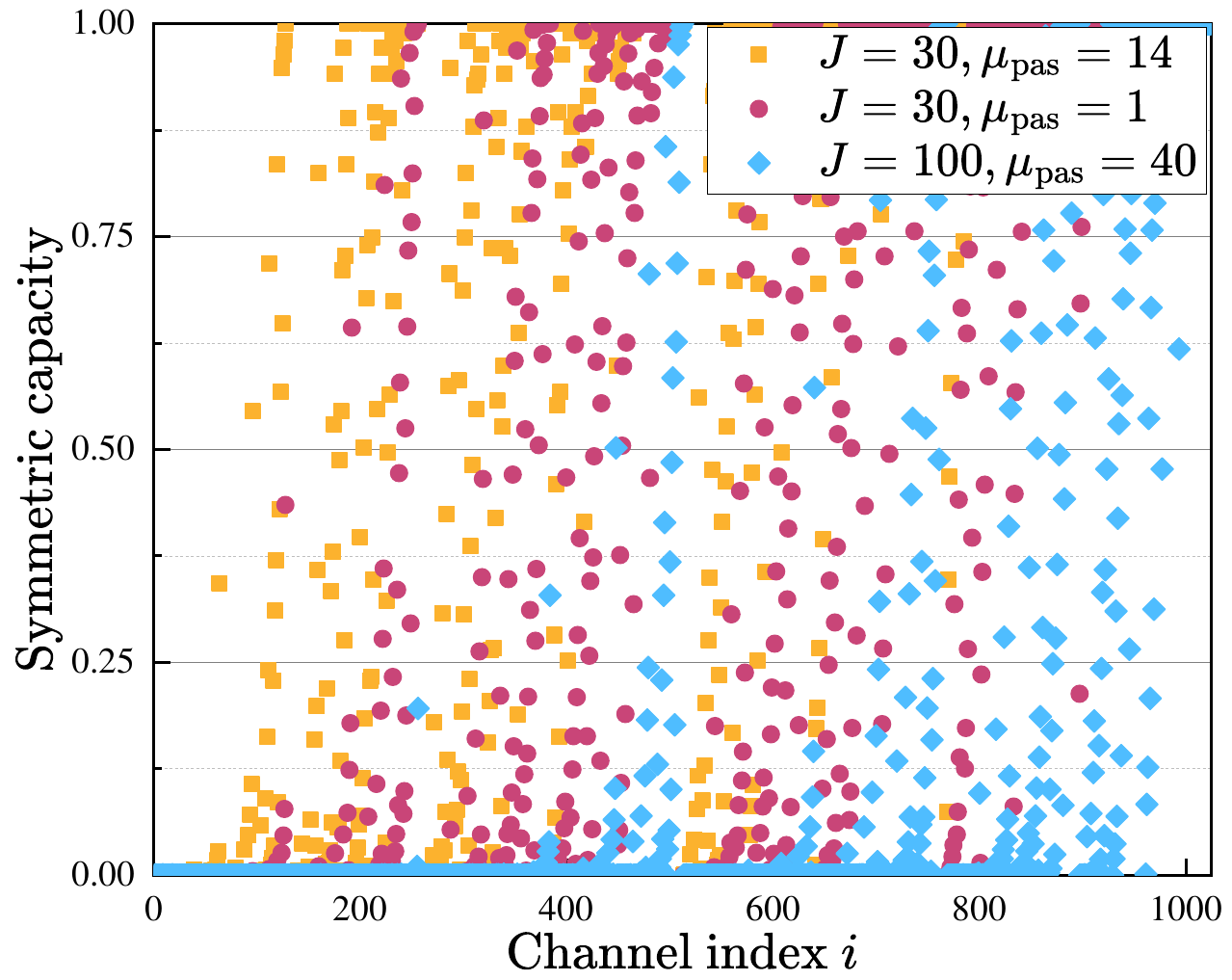}
    \caption{Capacities of individual bit channels for \( N = 1024 \), with the following configurations: (1) \( J = 30 \), \( \mu_{\mathrm{pas}} = 1 \); (2) \( J = 30 \), \( \mu_{\mathrm{pas}} = 14 \); and (3) \( J = 100 \), \( \mu_{\mathrm{pas}} = 14 \).}
    \label{fig:bit channel}
\end{figure}

\begin{example}
    The system parameters are configured with block length $m = 1024$ and $\mathrm{SNR} = 5~\mathrm{dB}$. Three simulation scenarios are specified: 
    (1) $J = 30$, $\mu_{\mathrm{pas}} = 1$; 
    (2) $J = 30$, $\mu_{\mathrm{pas}} = 14$; 
    (3) $J = 100$, $\mu_{\mathrm{pas}} = 14$.
    
    Figure~\ref{fig:bit channel} shows the capacities of individual bit channels. For a fixed user count, larger $\mu_{\mathrm{pas}}$ yield stronger channel polarization effects. When both $J=30$ and $J=100$ utilize their respective optimal $\mu_{\mathrm{pas}}$ values, the polarization effect for $J=30$ substantially surpasses that for $J=100$.
    
    This further reveals that polarization channels with same indices exhibit different capacities under varying $J$ and $\mu_{\mathrm{pas}}$ configurations. Consequently, distinct systems require different construction schemes to identify their optimal polarized channels.    
    $\blacktriangle \blacktriangle$
\end{example}
By calculating the channel capacities of different polarized subchannels, we identify the indices in the set \( \mathcal{A} \), where subchannels with higher capacities are selected for inclusion in \( \mathcal{A} \). It is important to note that the construction scheme for the systematic polarized EP code adheres to the partial order \cite{partial_order}, and thus, the two-step encoding method is still employed to encode the systematic polarized EP code \cite{fast-scl}.

\section{Decoding of PA-FFMA systems}\label{decode}
In the FFMA system, the SCL decoding algorithm performs well when the number of information bits \( K \) and the number of parity bits \( R \) are comparable for each user. However, when \( K \) is small, the SCL algorithm becomes less effective, and in such cases, the BMD algorithm provides better performance. Therefore, this section introduces both the SCL and BMD decoding algorithms.

\subsection{SCL Decoding}
The SCL decoding algorithm can be used to decode the FFMA system because the superimposed signals of \( J \)-users, referred to as the CFSP blcok, form a codeword of the polarized EP code. Therefore, by obtaining all the LLRs, we can begin decoding the CFSP block ${\bf r}$, using SCL decoding.

The LLRs of the information symbols are given by
\[
l(y_i) = \frac{2y_i \sqrt{\mu_{\mathrm{inf}} P_{\mathrm{avg}}}}{\sigma^2},
\]
for \( 0 \le i < M \). The LLRs of the parity symbols, \( l(y_i) \) for \( M \le i < m \), are given by Eq. (\ref{LLR}).

The SCL decoding algorithm maintains \( L \) paths and scores each path with a \textit{path metric (PM)}, approaching maximum-likelihood decoding \cite{scl}. We employ lazy copying to reduce memory transfers, and compute the \( f \)-function and PM using the method described in \cite{llr-scl}.

The SCL decoder for the FFMA system shares the same complexity as a standard SCL decoder for a polar code, meaning its complexity is independent of the number of users. The only additional computational cost comes from calculating Eq. (\ref{LLR}).
For the information section, the LLRs require \( m \) multiplications. Each LLR of the parity symbol involves \( J \) additions and one multiplication, resulting in a total cost of \( JR \) additions and \( R \) multiplications for all parity bits.
With lazy copying, the SCL decoding stage incurs a cost of \( O(Lm \log_2 m) \). Assuming constant costs for additions and multiplications, the overall complexity is
\(
O(JR + Lm \log_2 m).
\)

\subsection{Top$L$-BMD Decoding}
In \cite{FFMA}, a BMD decoding algorithm for the PA-FFMA system is proposed. 
The BMD algorithm identifies the most reliable information bits to determine the most probable \( L \) candidates of user information blocks. After reconstructing and modulating each group of information bits, an estimated signal is formed. The minimum distance between the received signal and the estimated signal is then computed, yielding a suboptimal result.
Thus, the BMD algorithm serves as an approximation to the \textit{maximum a posteriori (MAP)} algorithm. Its core idea is to explore potential decoding outcomes within a predefined range, considering the parity bits in conjunction with the information bits, and evaluating the decoding accuracy by comparing the Euclidean distance to the received signal.

In this paper, we propose a Top$L$-BMD decoding algorithm. As defined in \cite{FFMA2}, the proposed Top$L$-BMD consists of two phases: the first phase is a search process to find possible user information block set using the Top$L$ algorithm, and the second phase involves minimum distance detection, aided by the parity section. The following subsections detail each phase.

\subsubsection{{\textbf{Phase I (Top$L$ searching algorithm)}}}
Phase I aims to identify the \( L \) most probable of candidate user information blocks according to the maximum-likelihood (ML) criterion, which is evaluated using the path matrix (PM). The necessary PM is provided in the following lemma.
\begin{lemma}
Let \( {\bf w} \) and \( \hat{\bf w} \) represent the {transmit and detected user information blocks}, respectively. The posterior probability of the estimate \( \hat w_0^{M-1} \) is given by:
\begin{equation}\label{probability_info}
    \begin{aligned}
        P({\bf w} &= \hat{w}_{0}^{M-1} \mid {\bf y} = y_0^{M-1}) =
        \frac{P({\bf w} = \hat{w}_{0}^{M-1}, {\bf y} = y_0^{M-1})}
             {P({\bf y} = y_0^{M-1})} \\[4pt]
        &= \prod_{i=0}^{M-1} \left( 1 +
           \bigl(e^{l(y_i)}\bigr)^{-(1 - 2\hat{w}_i)} \right)^{-1}.
    \end{aligned}
\end{equation}
Taking the natural logarithm of this expression as:
\begin{equation}
    \begin{aligned}
        \mathrm{PM} &= \sum_{j=0}^{M-1} \ln\left( 1 + \bigl(e^{l(y_j)}\bigr)^{-(1 - 2\hat{w}_j)} \right).
    \end{aligned}
\end{equation}
\end{lemma}

In the Top$L$ algorithm, LLRs are first hard-decided to form user block $\hat{\mathbf w}$. For any flip set, its PM is updated bit-by-bit, since flipping bit $i$ changes the PM by
\[
\Delta\mathrm{PM}_i
  =\ln(1+e^{l_i})-\ln(1+e^{-l_i})
  =l_i,
\]
with $l_i=|l(y_i)|$. As $l_i\ge0$, each flip increases the PM, so the hard-decision PM is minimal and can set to 0. Consequently, the PM of a flip set equals the sum of the $l_i$ values for its flipped bits.
A lower PM implies higher likelihood, and a min-heap efficiently selects the $L$ flip sets with smallest PMs without examining all $2^M$ possibilities. The output is
\(
\mathcal{F}=\{\mathcal{F}_0,\mathcal{F}_1,\dots,\mathcal{F}_{L-1}\},
\)
where each $\mathcal{F}_l$ lists the bit indices to flip and corresponds to the candidate decoding $\hat{\mathbf w}_l$.

Here, we present a description of the \texttt{Top$L$} algorithm, which is implemented using the min-heap data structure.

The min-heap, denoted by $\mathcal{H}$, stores pairs $(s,S)$, where $s$ is the sum of the elements in the flip set $S$.  
Each iteration proceeds as follows:
\begin{align}
    (s,S) &\gets \arg\min_{(s',S')\in\mathcal{H}} s', \label{mins}\\
    \mathcal{H} &\gets \mathcal{H}\setminus\{(s,S)\},\\
    \mathcal{F} &\gets \mathcal{F}\cup\{S\}. \label{F}
\end{align}
At the beginning of the first iteration, an empty flip set (with no flipped bits) is inserted into $\mathcal{H}$ as the initial element.

A min-heap is then used to select the top \(L\) subsets with the smallest PMs. Let the selected set of \(L\) user blocks be denoted as
\(
\Upsilon_{\rm inf} = \{\hat{\mathbf w}_1, \hat{\mathbf w}_2, \dots, \hat{\mathbf w}_L\},
\)
where each \(\hat{\mathbf w}_l\) represents a detected user block.

\begin{figure}[t]
    \centering
    \includegraphics[width=0.8\linewidth]{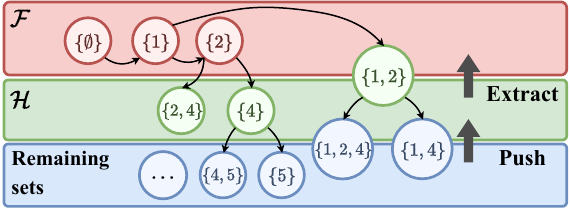}
    \caption{Dijkstra's algorithm on a directed binary tree. When Top$L$ reaches step 5, extract $\{1,2\}$ and generate $\{1,2,4\}$ and $\{1,4\}$ to the min-heap. 
    }
    \label{fig:tree}
\end{figure}

Next, set \(S\) grows along a binary tree:
\begin{equation}\label{Hadd1}
    \mathcal{H} \gets \mathcal{H} \cup \bigl\{(s + I_{i, 1},\, S \cup \{i\})\bigr\},
\end{equation}
\begin{equation}\label{Hadd2}
    \left\{
    \begin{aligned}
        &\Delta \gets I_{i, 1} - I_{i - 1, 1}, \\
        &\mathcal{H} \gets \mathcal{H} \cup
           \bigl\{(s + \Delta,\, (S \setminus \{i\}) \cup \{i + 1\})\bigr\}.
    \end{aligned}
    \right.
\end{equation}
Equation~\eqref{Hadd1} adds index \(i\), and \eqref{Hadd2} swaps it with \(i+1\). This is like running Dijkstra on a tree \cite{DIJ} (see Fig.~\ref{fig:tree}), visiting subsets in order without full enumeration.

Through the min-heap algorithm, we can identify the Top $L$ candidates with the minimum PMs.

\begin{example}\label{exampletree}
To obtain the five smallest subsets of $(1,2,4,5)$:

\begin{enumerate}
  \item Start with $\emptyset$ in $\mathcal{H}$.
  \item Extract $\emptyset$, append $1\rightarrow\{1\}$, push it, record $\emptyset$.
  \item Extract $\{1\}$, append $2\rightarrow\{1,2\}$, replace $1\rightarrow\{2\}$, push both, record $\{1\}$.
  \item Extract $\{2\}$, append $4\rightarrow\{2,4\}$, replace $2\rightarrow\{4\}$, push both, record $\{2\}$.
  \item Extract $\{1,2\}$, append $4\rightarrow\{1,2,4\}$, replace $2\rightarrow\{1,4\}$, record $\{1,2\}$.
  \item Extract $\{4\}$ and record it.
\end{enumerate}

The five subsets are $\emptyset$, $\{1\}$, $\{2\}$, $\{1,2\}$, and $\{4\}$. $\blacktriangle\blacktriangle$
\end{example}

The binary-tree view guarantees that every subset is reachable:
the left subtree of $\{1\}$ keeps $1$, the right discards it, and so on,
covering all $2^{JK}$ possibilities.

\begin{algorithm}[t]
    \caption{Top$L$}\label{SetTopL}
    \SetKwInOut{KwIn}{Input}
    \SetKwInOut{KwOut}{Output}
    \KwIn{Non-negative vector $\boldsymbol{l}$, list size $L$}
    \KwOut{Top-$L$ sets with smallest sums $\mathcal {F}$}
    $I \leftarrow \text{sort}\bigl(\{(l_i,i)\}_{i=0}^{n-1} \text{ by } l_i \bigr)$\;
    $\mathcal{H} \leftarrow \{(0,\varnothing)\},\;
      \mathcal {F} \leftarrow \{\varnothing\}$\;
    \While{$L>0$}{
        pop $(s,S)$ from $\mathcal{H}$ and add to $\mathcal {F}$ via Eqs.~(\ref{mins})–(\ref{F})\\
        $i \leftarrow \mathbf{1}_{S\neq\varnothing}\,(\max S+1)$\;
        \If{$i<|\boldsymbol{l}|$}{push by Eq.~(\ref{Hadd1})\;}
        \If{$0<i<|\boldsymbol{l}|$}{push by Eq.~(\ref{Hadd2})\;}
        $L \leftarrow L-1$\;
    }
    $\mathcal {F} \leftarrow \bigl\{\{I_{k,2}\mid k\in S\}\mid S\in\mathcal {F}\bigr\}$\;
    \Return{$\mathcal {F}$}
\end{algorithm}

\subsubsection{{\textbf{Phase II (minimum distance detection with partial re-encoding algorithm)}}}
In the second phase, codewords are generated from the detected user information block set $\Upsilon_{\rm inf}$. Phase II involves finding the minimum distance among the $L$ candidates to identify the closest optimal result \cite{FFMA}.

Let us consider the $l$-th candidate $\hat{\mathbf{w}}_l \in \Upsilon_{\rm inf}$ as an example to analyze the process. The procedure consists of three steps:
\begin{itemize}
    \item \textbf{Step 1:} Bit extraction. The bit sequences of $J$ users are extracted from $\hat{\mathbf{w}}_l$. For the $j$-th user, the bit sequence is given by
    \begin{equation}\label{newbj}
        \hat{\mathbf{b}}_{l,j} \gets \mathbf{\hat{w}}_{l,(j-1)K}^{\,jK}.
    \end{equation}

    \item \textbf{Step 2:} EP encoding. A CRC-aided EP encoder is applied to the bit sequence $\hat{\mathbf{b}}_{l,j}$ of the $j$-th user, yielding the encoded sequence
    \begin{equation}\label{newcj}
        \hat{\mathbf{c}}_{l,j} \gets \text{EP\_Encoder}(\hat{\mathbf{b}}_{l,j},\, j,\, N,\, C_L).
    \end{equation}

    \item \textbf{Step 3:} Signal update. Based on the output element $\hat{\mathbf{c}}_j$ of the $j$-th user, the signal is updated via the ${\rm F_{F2C}}$ operation:
    \begin{equation}\label{newr}
        \hat{\mathbf{r}}_l \gets \hat{\mathbf{r}_l} + \sqrt{\mu_{\rm inf} P_{\rm avg}} \circ \bigl(1 - 2 \hat{\mathbf{c}}_{l,j}\bigr).
    \end{equation}
\end{itemize}

Given the set of \(L\) candidate user blocks \(\Upsilon_{\rm inf} = \{ \hat{\mathbf w}_1, \hat{\mathbf w}_2, \dots, \hat{\mathbf w}_L \}\), their corresponding CFSP blocks form the set
\(
\Omega_{\hat{\mathbf r}} = \{\hat{\mathbf r}_1, \dots, \hat{\mathbf r}_L\}.
\)
Each block in $\Upsilon_{\rm inf}$ corresponds to a block in $\Omega_{\hat{\mathbf r}}$ by the index $l$, i.e., $\hat{\bf w}_2 \Leftrightarrow  \hat{\mathbf r}_l$. The detected user information block $\hat{\bf w}_l$ is determined by the CFSP block with the smallest distance to the received signal:
\[
    l^{\star} = \arg\min_{1\le l\le L} \|\mathbf y - \hat{\mathbf r}_l\|_{1},            
\]
where $l^{\star}$ is the index.

The decoder can skip full re-encoding of each flip set. It first encodes only the hard-decision bits to form each user's base codeword, then adds the modulated signals. When a bit flips, the new codeword is simply the base codeword XOR the flip vector, so no extra re-encoding is needed. 
Each user updates their detected $\hat{\mathbf{r}}$ by removing the old part and adding the new one:
    \begin{align}
        \mathbf{c}_j' &= \Bigl(\bigoplus_{k \in g} \mathbf{\alpha}^{l_{k,1}}\Bigr) \boldsymbol{\oplus} \mathbf{c}_j \label{Cj} \\   
        \hat{\mathbf{r}}' &= \hat{\mathbf{r}} - \sqrt{\mu_{\rm inf} P_{\rm avg}} \bigl( 1 - 2 \mathbf{c}_j \bigr)
        + \sqrt{\mu_{\rm inf} P_{\rm avg}} \bigl( 1 - 2 \mathbf{c}_j' \bigr) \label{r'}.
\end{align}
where $\mathbf{c}_j$ represents the original codeword, and $\mathbf{c}_j'$ is the updated codeword for user $j$. 
We partition the bit indices in $\mathcal{F}_l$ into subsets $g$ according to their respective users, and denote the collection of all such subsets by $\mathcal{G}$.

Now, we present an example to anlayze the flip process.

\begin{example}\label{md}
  We revisit Example 3, where the EPs have already been assigned to two users. 
  Assume that the hard-decision result of the parallel user block is 
  \({\bf \hat w} = ({\bf b}_1, {\bf b}_2) = (1, 0, 0, 1)\). Therefore, the information sections for users 1 and 2 are respectively:
  \begin{equation*}
    \begin{array}{ll}
        {\bf c}_{1, \rm inf} = ({\bf b}_1, {\bf 0}) = (\textcolor{blue}{1, 0}, 0, 0), \\
        {\bf c}_{2, \rm inf} = ({\bf 0}, {\bf b}_2) = (0, 0, \textcolor{blue}{0, 1}).
    \end{array}
  \end{equation*}
  Next, by using the CRC-aided systematic EP encoder, we obtain the output elements for users 1 and 2 as:
  \begin{equation*}
    \begin{array}{ll}
        {\bf c}_1 = (1, 0, 0, 0, 1, 1, 1, 0), \quad
        {\bf c}_2 = (0, 0, 0, 1, 0, 1, 1, 1).
    \end{array}
  \end{equation*}
  The corresponding modulated signals are:
  \begin{equation*}
    \begin{array}{ll}
        {\bf x}_1 = (-1, +1, \textcolor{blue}{0}, \textcolor{blue}{0}, -1, -1, -1, +1), \\
        {\bf x}_2 = (\textcolor{blue}{0}, \textcolor{blue}{0}, +1, -1, +1, -1, -1, -1).
    \end{array}
  \end{equation*}
  The regenerated CFSP block is:
  \[
      \hat{{\bf r}}_1 = {\bf x}_1 + {\bf x}_2 
             = (-1, +1, +1, -1, 0, -2, -2, 0).
  \]
  
  Now assume that the bit \(\hat w_1\) flips, which means \({\bf \hat w} = (1, \textcolor{red}{1}, 0, 1)\). Then, user 1's information section becomes ${\bf c}_{1, \rm inf} = (\textcolor{red}{1, 1}, 0, 0)$, and the output element is:
  \[
      {\bf c}_1 = (\textcolor{red}{1, 1}, \textcolor{blue}{0}, \textcolor{blue}{0}, 0, 0, 1, 1),
  \]
  with the corresponding modulated signal:
  \[
      {\bf x}'_1 = (\textcolor{red}{-1, -1}, \textcolor{blue}{0}, \textcolor{blue}{0}, +1, +1, -1, -1).
  \]
  The updated CFSP block is:
  \begin{align*}
      \hat{{\bf r}_2} = \hat{{\bf r}_1} - {\bf x}_1 + {\bf x}'_1 
        = (-1, -1, +1, -1, +2, 0, -2, -2).
  \end{align*}
  The Top $L$ detected CFSP set is:
  \(
      \Omega_{\hat{\bf r}} = \{\hat{{\bf r}_1}, \hat{{\bf r}_2}\}.
  \)
  $\blacktriangle\blacktriangle$
\end{example}

\begin{algorithm}[t]
    \caption{Top$L$-BMD Decoding}\label{BMDDecoding}
    \SetKwInOut{KwIn}{Input}
    \SetKwInOut{KwOut}{Output}
    \KwIn{Received signal $\mathbf{y}$, list size $L$, 
    users' number $J$, bits per user $K$, 
    CRC length $C_L$, 
    generator matrix $\mathbf{G}_{\rm M, sym, crc}$}
    \KwOut{Decoded bits ${\mathbf{w}}$}
  
    $\boldsymbol\ell\!\leftarrow\!\dfrac{2\sqrt{\mu_{\rm inf}P_{\rm avg}}}{\sigma^{2}}\,y_0^{JK-1}$,
    $\mathcal F\!\leftarrow\!\text{Top$L$}(|\boldsymbol\ell|,L)$\\
    $\hat{\mathbf w}\!\leftarrow\!\mathbf 1_{\{\boldsymbol\ell<0\}},
     \mathbf r\!\leftarrow\!\mathbf 0_{1\times m}$, $\mathbf b\gets\mathbf 0_{J\times K}$, ${\mathbf c}\gets\mathbf 0_{J\times m}$\\
  
    \For{$j=1$ {\bf to} $J$}{update $(\mathbf b_j,\mathbf c_j,\mathbf r)$ via \eqref{newbj}--\eqref{newr}}
  
    $\delta\!\leftarrow\!\|\mathbf r-\mathbf y\|_1$, $l^\star\!\leftarrow\!0$\\
  
    \For{$l=0$ {\bf to} $L-1$}{
        Group bit indices in $\mathcal{F}_l$ by user into subsets $\mathcal{G}$\\
        \For{$g\!\in\!\mathcal{G}$}
            {update $\hat{\mathbf r}$ via \eqref{Cj} and \eqref{r'} with $j \leftarrow \Bigl\lfloor\frac{g_0}{K}\Bigr\rfloor + 1$}    
            {$\delta\leftarrow\|\mathbf y-\hat{\mathbf r}_l\|_1,\,l^{\star} \leftarrow l$, if $\|\mathbf y-\hat{\mathbf r}_l\|_1<\delta$}                
    }
    \For{$i\in\mathcal F_{l^\star}$}{$w_i\!\leftarrow\!\hat w_i\oplus1$}
    \Return{$\mathbf w$}
  \end{algorithm}\label{GroupByUser}
  
\subsubsection{Complexity of the Top$L$-BMD Algorithm}
Algorithm~\ref{BMDDecoding} presents the Top-\(L\)-BMD decoder, which utilizes the CRC-aided systematic generator in~(\ref{G_syspolar}). Suppse that additions and multiplications have constant cost, the Top-\(L\)-BMD algorithm consists of two phases: the Top-\(L\) phase and the BMD phase.

In the Top$L$ phase, we do $L$ loops. Each loop pops and pushes one item in a heap of size up to $L$, which costs $O(\log L)$. Copying a flip set costs $O(n_{\max})$, where $n_{\max}$ is its largest size. Sorting $M$ likelihoods costs $O(M\log M)$. So the Top$L$ step costs
\(
O\bigl(L\log L + L\,n_{\max} + M\log M\bigr),
\)
and $n_{\max}\le\log L$.

For the BMD phase, the algorithm rebuilds $L$ candidate codewords by adding flipped rows in ${\text{GF}(2)}$, which requires $O(L n_{\max} R)$ operations. Forming each candidate signal and calculating its distance to the received vector costs $O(Lm)$. Hence, the overall complexity of the proposed Top$L$-BMD decoder is:
\[
O\!\left(L\log_2 L + L n_{\max} R + M\log_2(M) + Lm\right),
\]
which is primarily influenced by the parameters $L$, $M$ and $m$.
    
\section{Simulation Results}\label{numeric}
In this section, we present simulations to validate the proposed systems. First, we analyze the capacity of a GMAC through simulations. Next, we perform Monte Carlo simulations to determine the optimal power allocation that maximizes the capacity. Finally, we assess the bit error rate (BER) performance of our FFMA systems in comparison with classical polar spreading systems.

\subsection{Channel Capacity and Power Allocation} 

\begin{figure}[t]
    \centering
    \includegraphics[width=0.8\linewidth]{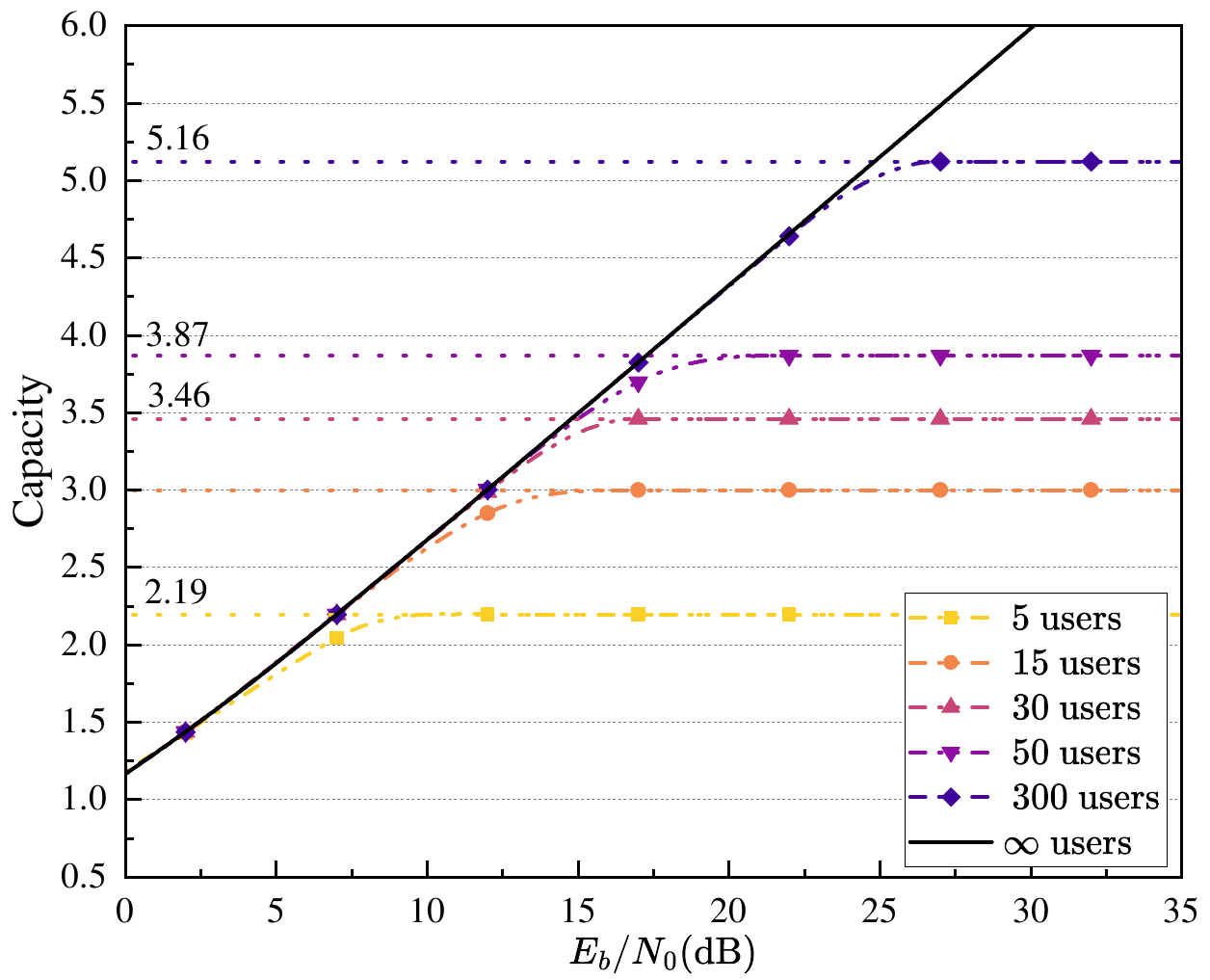}
    \caption{The capacity of a GMAC, where the number of users are $J$ = $5$, $15$, $30$, $50$, and $300$, respectively.}
    \label{fig:MI}
\end{figure}

The channel capacity of the GMACs is shown in Fig.~\ref{fig:MI}. As \( E_b/N_0 \) increases, the channel capacity gradually approaches the entropy \( {\rm H}({\mathcal R}) \) of the CFSP variable \( \mathcal R \), where \( \mathcal R \in \Omega_r \). As the number of users \( J \) increases, the entropy grows approximately logarithmically with respect to \( J \), i.e.,
\begin{equation*}
    {\rm H}({\mathcal R}) = \sum_{\nu = 0}^{J} {\mathcal P}_r(\nu) \log \frac{1}{{\mathcal P}_r(\nu)} < \log_2 (J+1),
\end{equation*}
which grows slower than the linear increase of \( J \). This implies that for a given \( E_b/N_0 \), the channel capacity of a \( J \)-user system is generally smaller than the sum of the individual channel capacities of \( J \) users.
However, using the GMAC channel gives each user more parity bits to use, enabling greater coding gain. If this coding gain outweighs the slow logarithmic growth in channel capacity, overall performance improves.

\begin{figure}[t]
    \centering
    \includegraphics[width=0.8\linewidth]{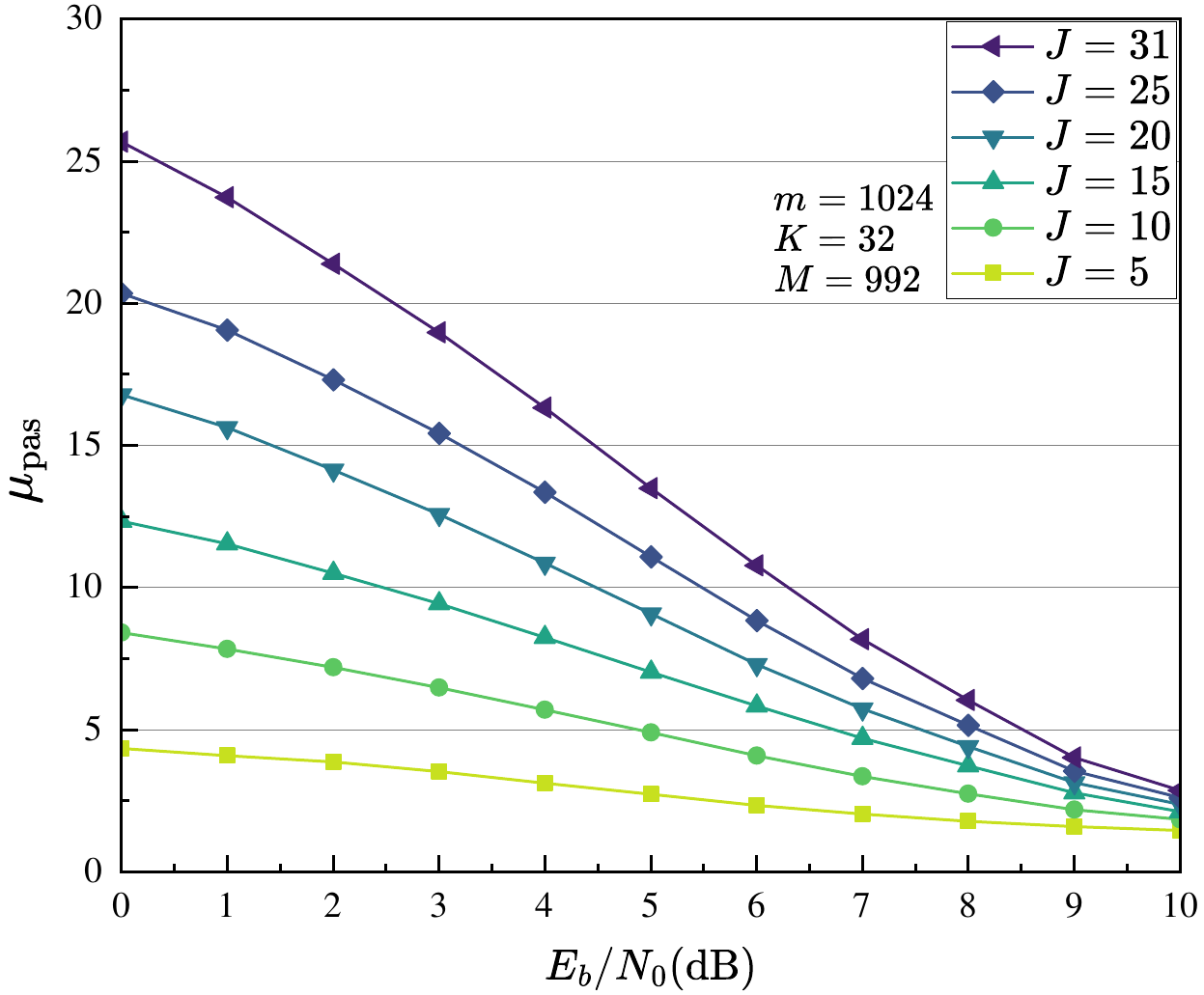}
    \caption{Polarization ajusted scaling factor $\mu_{\mathrm{pas}}$ for maximizing the channel capacity, where $J = 5, 10, 15, 25, 31$.}
    \label{fig:power_dis}
\end{figure}

Fig. \ref{fig:power_dis} shows the power allocation results for the proposed PA-FFMA systems based on Monte Carlo simulations. The x-axis represents \( E_b/N_0 \), while the y-axis corresponds to the power-adjusted scaling factor \( \mu_{\rm pas} \). In the simulation, the system parameters are set as follows: the number of degrees of freedom (DoFs) is \( m = 1024 \), the number of bits per user is \( K = 32 \), the number of EPs is \( M = 992 \), and the number of users varies as \( J = 5, 10, 15, 20, 25, 31 \). Each curve corresponds to a specific user count, illustrating the optimal power allocation scheme under different \( E_b/N_0 \) conditions.
For a fixed user count, \( \mu_{\rm pas} \) decreases with increasing \( E_b/N_0 \), indicating a tendency to allocate more power to the parity section. This happens because the channel capacity of the parity section exceeds that of the information section under favorable channel conditions.
For a fixed \( E_b/N_0 \), as the user count increases, \( \mu_{\rm pas} \) further decreases, signifying greater power allocation to the information section due to multi-user superposition.

\subsection{BER performance}
\begin{figure}[t]
    \centering
    \includegraphics[width=0.8\linewidth]{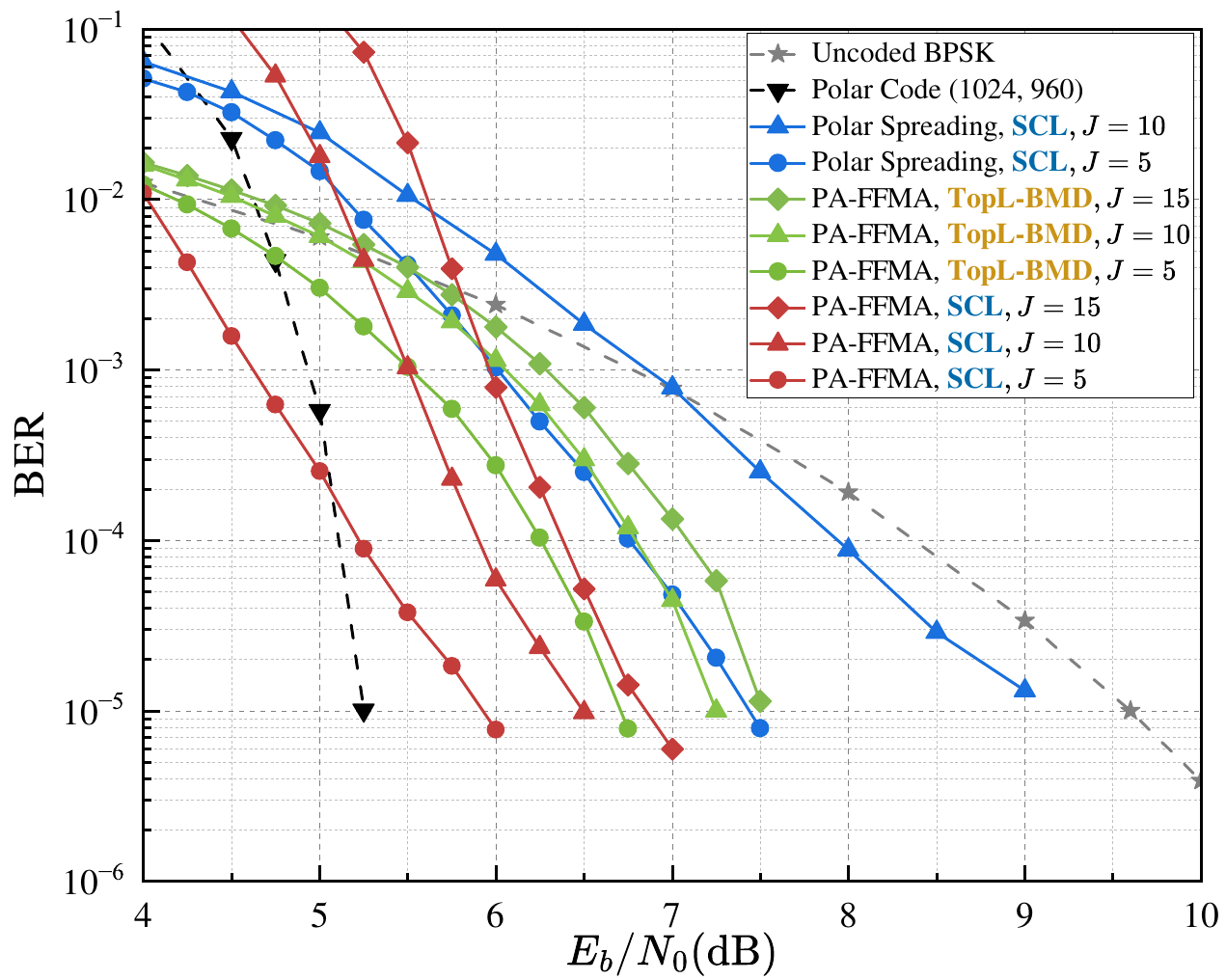}
    \caption{Error performance comparison between PA-FFMA and polar spreading systems, where $K=64$, and $J = 5, 10, 15$.}
    \label{fig:m=1024_J=15K64}
\end{figure}
  
\begin{figure}[t]
\centering
\includegraphics[width=0.8\linewidth]{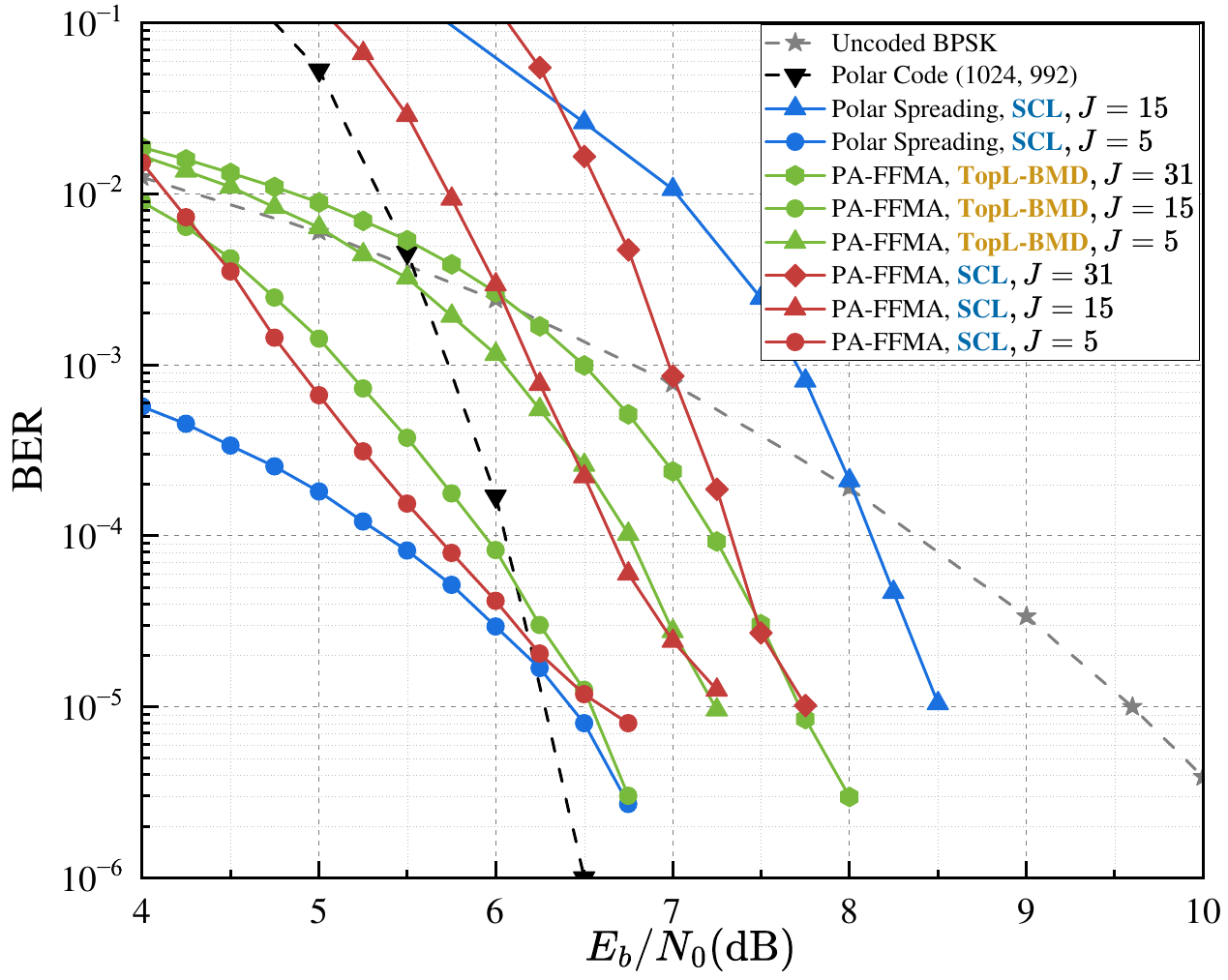}
\caption{Error performance comparison between PA-FFMA and polar spreading systems, where $K=32$, and $J = 5, 15, 31$.}
\label{fig:m=1024_J=31K32}
\end{figure}

Fig.~\ref{fig:m=1024_J=15K64} shows the BER performance of our proposed PA-FFMA systems compared to the state-of-the-art polar coding with random spreading scheme~\cite{PS}. The proposed PA-FFMA system is configured with the following parameters: the number of DoFs \(m=1024\), the number of EPs \(M=960\), the number of bits per user \(K=64\), the CRC length \(\mathrm{CRC}=8\), and it employs two decoding algorithm:SCL and Top$L$-BMD with \(L=512\) decoding path, and polarization ajusted scaling factor explores the optimal \(\mu_{\rm pas}\).

The baseline system utilizes 5G-standard polar codes with parameters \( (128, 64, \mathrm{CRC} = 8) \), along with length-$8$ random Gaussian spreading sequences (assumed to be known at the receiver), and is decoded using an iterative, convergence-guaranteed algorithm. In each iteration, MMSE detection is first applied to obtain the LLRs, followed by SCL decoding. The decoded codewords that pass the CRC check are subtracted using successive interference cancellation, and the remaining signals are reprocessed by MMSE detection. This process is repeated until all users pass the CRC check or no further decoding is possible. All simulations assume a total transmit power of \( m P_{avg} = 1024 P_{avg} \), with randomly generated codewords.

First, we analyze the decoding algorithms in the polarized EP-based FFMA system. The results demonstrate that the SCL decoder consistently outperforms the Top$L$-BMD decoder across all tested user loads. At light user load, e.g., $J = 5$, the SCL decoder achieves a BER of $10^{-5}$ at $6$~dB, whereas the Top$L$-BMD decoder requires nearly $6.75$~dB to reach the same BER level. This indicates a performance gain of approximately $0.75$~dB for the SCL algorithm. The performance gap between the two decoders narrows with increasing user load. Notably, when $J = 15$, the gain reduces to nearly 0.5~dB.

Then, we compare our proposed FFMA system with the SCL decoding algorithm to the polar random spreading system. It is evident that as the number of users increases, our proposed FFMA system exhibits significantly better error performance compared to the polar random spreading system. For instance, for $J = 5$ users and a BER of $10^{-5}$, our proposed FFMA system achieves a gain of approximately {$1.5$ dB} over the polar random spreading system. For $J = 10$ users and BER = $10^{-5}$, the gain increases to around {$2.5$} dB. When $J = 15$, our proposed FFMA system still functions and provides a gain of about {$2.85$} dB compared to the uncoded case, whereas the polar random spreading system fails to perform at this point. This demonstrates the superiority of our proposed polarized EP-based FFMA system in a GMAC.

Next, we examine the impact of a small payload size, $K = 32$, which is half of the payload in Fig.~\ref{fig:m=1024_J=31K32}. The proposed polarized EP-based FFMA system utilizes polar codes with the following parameters: the number of DoFs $m = 1024$, the number of EPs $M = 992$, the number of bits per user $K = 32$, and a CRC length of $\mathrm{CRC} = 8$. In contrast, the baseline system employs polar codes with parameters $(64, 32, \text{CRC} = 8)$, combined with length-16 near-orthogonal random spreading sequences.

For the $K=32$ case, we observe that the Top$L$-BMD decoding outperforms the SCL decoding in terms of error performance for the same user count. For instance, when there are $J=5$ users, the SCL decoder achieves a BER of $10^{-5}$ at approximately $6.75$ dB, while the Top$L$-BMD decoder reaches a BER of $10^{-5}$ at $6.5$ dB, yielding a $0.25$ dB improvement. This contrasts with the result in Fig.~\ref{fig:m=1024_J=31K32}, where the SCL decoder outperforms the Top$L$-BMD decoder. As the user count increases, the SCL decoder continues to lag behind the Top$L$-BMD decoder, although their performance converges at high $E_b/N_0$.

Next, we compare our proposed FFMA system with the Top$L$-BMD decoding algorithm against the polar random spreading system. As with the $K = 64$ case discussed earlier, as the number of users increases, our proposed FFMA system demonstrates significantly better error performance compared to the polar random spreading system. For example, for $J = 15$ users and a BER of $10^{-5}$, our proposed FFMA system achieves a gain of approximately {$1.25$} dB over the polar random spreading system. When $J = 31$, our proposed FFMA system continues to function and provides a gain of about {$1.85$} dB compared to the uncoded case, while the polar random spreading system fails to perform at this point. These results are consistent with those observed in the $K = 64$ case.

\begin{table}[t]
    \centering
    \caption{Decoding Complexity Comparison.}
    \label{tab:complexity_aligned}
    \footnotesize
    \renewcommand{\arraystretch}{1.2} 
    \begin{tabularx}{\columnwidth}{@{}l>{\raggedright\arraybackslash}X@{}}
        \toprule
        \textbf{Algorithm} & \textbf{Complexity} \\
        \midrule
        FFMA-SCL        & $O(JR + Lm\log m)$ \\
        FFMA-Top$L$-BMD   & $O\big(L\log L + Ln_{\max}R + JK\log(JK) + Lm\big)$ \\
        Polar Spreading & $O\big(T(n_s^3 + Jn_s(K+R) + JL(K+R)\log (K+R))\big)$ \\
        \bottomrule
    \end{tabularx} 
\end{table}


Furthermore, we compare the decoding complexity of the FFMA and polar random spreading systems, as summarized in Table I. In this table, $n_s$ denotes the spreading length, and $T$ represents the number of SIC iterations. From Table I, we observe the following three key points:

     FFMA-SCL's complexity term \(L m \log m\) is usually well below the polar-spreading term \(T J (K+R)\log(K+R)\).
    For instance, with \(m\!=\!1024,\;K\!=\!R\!=\!64,\;J\!=\!15,\;T\!=\!3\):
    \(
       T J (K+R)\log(K+R)\approx4.0\times10^{4}
       \;>\;
       m \log m \approx 1.0\times10^{4},
    \)
    so FFMA-SCL is clearly faster.

     In FFMA-Top\(L\)-BMD, the minor terms \(L\log L\) and \(L n_{\max} R\) are negligible.  
    The leading costs \(JK\log(JK)\) and \(Lm\) are below the polar-spreading burdens  
    \(T J n_s (K+R)\) and \(T J L (K+R)\log(K+R)\), resulting in faster decoding.

     Considering only the dominant term in the complexity, the FFMA-SCL scales as $L m \log m$, while the FFMA-Top$L$-BMD scales as $L m$. Thus, the Top$L$-BMD algorithm exhibits lower complexity. Moreover, in simulations, the Top$L$-BMD decoder operates approximately six to seven times faster than the SCL decoder.

This complexity analysis confirms the dual advantages of FFMA: superior error-rate performance combined with lower computational demands compared to polar spreading systems.

\section{CONCLUSION}\label{conclusion}

In this paper, we introduce polarized EP codes for FFMA systems, including their encoding, construction, and decoding processes. We begin by presenting the basic definition of polarized EP codes, followed by the CRC-aided systematic polarized EP code, which partitions the FFMA system into an information section and a parity section. This partition allows for the reallocation of unused power between these two sections, with different polarization-adjusted factors. 
Next, we analyze the channel capacity of the proposed system, where the GMAC consists of two concatenated channels: one being BI-ASC and the other MI-NSC. We also propose two decoding algorithms for the PA-FFMA systems. For users with relatively large payloads, the proposed SCL algorithm delivers superior error performance. In contrast, the Top$L$-BMD decoding algorithm offers better performance when the payload is small.
When compared to the state-of-the-art polar code with a random spreading scheme, the proposed PA-FFMA system demonstrates well-behaved error performance.
\bibliographystyle{IEEEtran}
\bibliography{references}

\end{document}